%% file: main.tex
\documentclass[twocolumn,trackchanges,floatfix]{aastex631}
\usepackage{amsmath}
\usepackage{longtable}
\usepackage{multirow}

\newcommand{\AGAMA}{\texttt{AGAMA}}
\newcommand{\MWMMXVII}{\texttt{MW2017}}
\newcommand{\HDBSCAN}{\texttt{HDBSCAN}}
\newcommand{\peri}{\text{peri}}
\newcommand{\apo}{\text{apo}}
\newcommand{\maxtext}{\text{max}}

\shorttitle{BNB DTGs}
\shortauthors{Shank et al.}

\begin{document}

\title{Dynamically Tagged Groups of Metal-Poor Stars from the Best \& Brightest Survey}

\correspondingauthor{Derek Shank}
\email{dshank1@nd.edu}

\received{September 17, 2021}
\revised{November 9, 2021}
\accepted{December 1, 2021 in The Astrophysical Journal}

\author[0000-0001-9723-6121]{Derek Shank}
\affiliation{Department of Physics, University of Notre Dame, Notre Dame, IN 46556, USA}
\affiliation{Joint Institute for Nuclear Astrophysics -- Center for the Evolution of the Elements (JINA-CEE), USA}

\author[0000-0003-4573-6233]{Timothy C. Beers}
\affiliation{Department of Physics, University of Notre Dame, Notre Dame, IN 46556, USA}
\affiliation{Joint Institute for Nuclear Astrophysics -- Center for the Evolution of the Elements (JINA-CEE), USA}

\author[0000-0003-4479-1265]{Vinicius M. Placco}
\affiliation{NSF’s NOIRLab, 950 N. Cherry Ave., Tucson, AZ 85719, USA}

\author[0000-0002-9269-8287]{Guilherme Limberg}
\affiliation{Universidade de S\~{a}o Paulo, Instituto de Astronomia, Geof\'{i}sica e Ci\^{e}ncias Atmosf\'{e}ricas, Departamento de Astronomia, SP 05508-090, S\~{a}o Paulo, Brazil}

\author{Emma Jaques}
\affiliation{Department of Physics, University of Notre Dame, Notre Dame, IN 46556, USA}

\author[0000-0002-8129-5415]{Zhen Yuan}
\affiliation{Universit\'e de Strasbourg, CNRS, Observatoire Astronomique de Strasbourg, UMR 7550, F-67000 Strasbourg, France}

\author[0000-0001-5761-6779]{Kevin C.\ Schlaufman}
\affiliation{Department of Physics and Astronomy, Johns Hopkins University, 3400 N Charles Street, Baltimore, MD 21218, USA}

\author[0000-0003-0174-0564]{Andrew R. Casey}
\affiliation{School of Physics $\&$ Astronomy, Monash University, Wellington Road, Clayton 3800, Victoria, Australia}
\affiliation{ARC Centre of Excellence for All Sky Astrophysics in 3 Dimensions (ASTRO 3D), Canberra, ACT 2611, Australia}

\author[0000-0003-3250-2876]{Yang Huang}
\affiliation{South-Western Institute for Astronomy Research, Yunnan University, Kunming 650500, People's Republic of China}

\author[0000-0003-0852-9606]{Young Sun Lee}
\affiliation{Department of Astronomy and Space Science, Chungnam National University, Daejeon 34134, Republic of Korea}

\author[0000-0001-6924-8862]{Kohei Hattori}
\affiliation{National Astronomical Observatory of Japan, 2-21-1 Osawa, Mitaka, Tokyo 181-8588, Japan}
\affiliation{Institute of Statistical Mathematics, 10-3 Midoricho, Tachikawa, Tokyo 190-0014, Japan}

\author[0000-0002-7529-1442]{Rafael M. Santucci}
\affiliation{Universidade Federal de Goi\'as, Instituto de Estudos Socioambientais, Planet\'ario, Goi\^ania, GO 74055-140, Brazil}
\affiliation{Universidade Federal de Goi\'as, Campus Samambaia, Instituto de F\'isica, Goi\^ania, GO 74001-970, Brazil}

\date{\today}

\begin{abstract}
    Orbital characteristics based on Gaia Early Data Release 3 astrometric parameters are analyzed for ${\sim} 4000$ metal-poor stars ([Fe/H] $\leq -0.8$) compiled from the Best $\&$ Brightest survey. Selected as metal-poor candidates based on broadband near- and far-IR photometry, $43\%$ of these stars had medium-resolution ($1200 \lesssim R \lesssim 2000$) validation spectra obtained over a seven-year campaign from $2014$ to $2020$ with a variety of telescopes. The remaining stars were chosen based on photometric metallicity determinations from the Huang et al. recalibration of the Sky Mapper Southern Survey. Dynamical clusters of these stars are obtained from the orbital energy and cylindrical actions using the \HDBSCAN ~unsupervised learning algorithm. We identify $52$ Dynamically Tagged Groups (DTGs) with between $5$ and $21$ members; $18$ DTGs have at least $10$ member stars. Milky Way (MW) substructures such as Gaia-Sausage-Enceladus, the Metal-Weak Thick-Disk, Thamnos, the Splashed Disk, and the Helmi Stream are identified. Associations with MW globular clusters are determined for $8$ DTGs; no recognized MW dwarf galaxies were associated with any of our DTGs. Previously identified dynamical groups are also associated with our DTGs, with emphasis placed on their structural determination and possible new identifications. Chemically peculiar stars are identified as members of several DTGs, with $6$ DTGs that are associated with \textit{r}-process-enhanced stars. We demonstrate that the mean carbon and $\alpha$-element abundances of our DTGs are correlated with their mean metallicity in an understandable manner.  Similarly, we find that the mean metallicity, carbon, and $\alpha$-element abundances are separable into different regions of the mean rotational-velocity space. 
\end{abstract}


\section{Introduction}\label{sec:Introduction}

The identification of substructure in the halo of the Galaxy has been underway for at least three decades, essentially as soon as moderate to large samples of stellar halo tracers became available.  For example, \citet{Sommer-Larsen1987} noted the possible presence of a physical group of field horizontal-branch (FHB) stars in a limited sample of blue objects on a single wide-field Schmidt plate; the proposed group contained five stars with a small spread in distance moduli ($\sigma$ = 0.16 mag) and radial velocity dispersion $\sigma \leq 20$ km s$^{-1}$.  This finding motivated a search for other such objects within a catalog of some 200 FHB stars with available photometric and spectroscopic measurements by \citet{Beers1987}, who found a similar group in the direction toward the Galactic Bulge comprising at least five members (see Table 1 of \citealt{Doinidis1989}).  The much larger catalog of 4400 candidate FHB stars from \citet{Beers1988} enabled \citet{Doinidis1989} to carry out a statistical analysis of the two-point correlation function for stellar pairs of FHB candidates.  Their finding that the catalog contains a significant excess of pairs with separations $\theta < 10\arcmin$ (corresponding to physical separations $r \leq 25$ pc) at distances 5-8 kpc from the Sun strongly indicated the ubiquitous presence of such structures throughout the inner halo of the Galaxy; they suggested that these might be tracers of recently disrupted Galactic satellites or globular clusters.

Rapid progress in the identification of substructures followed (including numerous streams and over-densities), as sample sizes expanded, in particular with the advent of the Sloan Digital Sky Survey (SDSS; \citealt{York2000}); see \citet{Helmi2008} and \citet{Belokurov2013} for a review of this early work.   

The known tidal streams that are detectable as surface-brightness enhancements in the halo represent only the most recent infall events that have not yet been phase-mixed. In order to create a more complete description of the merger history of the Milky Way (MW), we must find debris from earlier accretion events. Conservation of phase-space density in collisionless stellar systems implies that, as tidal-debris structures become spatially dispersed, they must become kinematically colder (for a limited time – after many crossing times the debris will be completely phase-mixed with the halo). Thus, some substructures that are no longer visible as spatial over-densities but that have not been completely phase-mixed can be detected as kinematically cold peaks relative to ``smooth” halo populations. 

\citet{Schlaufman2009} pioneered a technique to search for ``Elements of Cold Halo Substructure” (ECHOS) in SDSS data. These authors performed a separate search in each of 137 individual SDSS ``plug-plates” covering $\sim 7$ square degrees on the sky, and identified statistically significant peaks among the radial velocities of stars selected to be in the inner halo, at distances of $10$ - $17.5$ kpc from the Sun. These authors estimated that roughly one-third of the metal-poor main-sequence turnoff stars in the inner halo are part of ECHOS. In a second paper, \citet{Schlaufman2011} explored the metallicities and $\alpha$-element abundances of the previously identified ECHOS, finding that these structures exhibit unique abundance signatures compared to inner-halo stars along the same lines-of-sight. Specifically, the ECHOS are, on average, more metal-rich ([Fe/H]\footnote{The standard definition for an abundance ratio of an element in a star $(\star)$ compared to the Sun $(\odot)$ is given by $[A/B] = (\log{N_{A}/N_{B}})_{\star} - (\log{N_{A}/N_{B}})_{\odot}$, where $N_{A}$ and $N_{B}$ are the number densities of atoms for elements $A$ and $B$.} $\sim -1.0$), and have lower $\alpha$-element abundances than typical halo stars, suggesting that ECHOS likely originated in relatively massive ($> 10^{9} M_{\odot}$) galaxies similar to the classical dwarf spheroidals.  They argued that the metallicities of ECHOS, combined with the stellar mass-metallicity relation for dwarf galaxies (e.g., \citealt{Kirby2013}), could be used to recreate the mass spectrum of the building blocks that contributed their stars to the Galactic halo.  Finally, \citet{Schlaufman2012} reported on an apparent transition from an inner halo that shows no spatial auto-correlation (as expected for in situ and/or early major-merger formation) to the outer regions of the halo ($> 15$ kpc) that exhibits strong auto-correlation on large scales (consistent with the outer halo being formed entirely from accreted satellites).  It is not difficult to see the connection between these prescient early studies and the later recognition of, e.g., the so-called Ancient Chronographic Sphere \citep{Santucci2015,Carollo2016} of FHB stars, and the Gaia-Sausage-Enceladus structures (GSE; \citealt{Belokurov2018,Haywood2018,Helmi2018}). 

When Galactic satellites are accreted and dispersed into the MW, the energies and dynamical actions of their member stars are expected to resemble those of their parent progenitor satellites \citep{Helmi1999}. The seminal work of \citet{Roederer2018} employed unsupervised clustering algorithms, an approach that has proven crucial to determine structures in the MW that are not revealed through large-scale statistical sampling methods. These authors were able to collect $35$ chemically peculiar (\textit{r}-process-enhanced) stars and determine their orbits.  Multiple clustering tools were applied to the orbital energy and actions to determine stars with similar orbital characteristics. This study revealed eight dynamical groupings comprising between two and four stars each. The small dispersion of each group's metallicity was noted, and explained as the possible origin of a satellite accretion event.

\citet{Yuan2020b} utilized the self-organizing map neural network routine StarGO \citep{Yuan2018} on the Large Sky Area Multi-Object Fiber Spectroscopic Telescope (LAMOST; \citealt{Cui2012}) Data Release 3 \citep{Li2018} stellar survey. These authors used StarGO to examine very metal-poor ([Fe/H] $\lesssim -1.8$) stars to seek dynamical clusters based on the derived energy, angular momentum, polar, and azimuthal (E,L,$\theta$,$\phi$) parameters. The authors identified $57$ dynamically tagged groups (DTGs), of which most are identified as part of GSE or Sequoia \citep{Myeong2019}, while $18$ are new structures. \citet{Limberg2021a} constructed DTGs from metal-poor stars in the HK \citep{Beers1985,Beers1992} and Hamburg/ESO \citep{Christlieb2008} surveys using the Hierarchical Density-Based Spatial Clustering of Applications with Noise (\HDBSCAN; \citealt{Campello2013}) algorithm over the orbital energy and cylindrical action space. The clustering procedure was able to identify 38 DTGs, with 10 of those being newly identified substructures. \citet{Gudin2021} extended the work by \citet{Roederer2018}, using a much larger sample of \textit{r}-process-enhanced stars (see their Table 1 for definitions). Also utilizing the \HDBSCAN ~algorithm, 30 Chemo-Dynamically Tagged Groups (CDTGs)\footnote{The distinction between CDTGS and DTGs is that the original stellar candidates of CDTGs are selected to be chemically peculiar in some fashion, while DTGs are selected from stars without detailed knowledge of their chemistry, other than [Fe/H].} were discovered. Their analysis revealed statistically significant similarities in the dispersions of stellar metallicity, carbon abundance, and \textit{r}-process-element abundances ([Sr/Fe],[Ba/Fe], and [Eu/Fe]), strongly suggesting these stars experienced similar chemical-evolution histories in their progenitor galaxies. 

This work aims to analyze the DTGs present in the Best and Brightest (B$\&$B; \citealt{Schlaufman2014}) survey, which contains metal-poor stellar candidates brighter than \textit{V} $\sim 14$. The association of our identified DTGs with recognized Galactic substructures, previously known DTGs/CDTGs, Globular Clusters, and Dwarf Galaxies is explored, with the most interesting stellar populations being noted for future high-resolution follow-up studies.

This paper is outlined as follows. Section \ref{sec:Data} describes the B$\&$B sample along with their associated astrometric parameters. The dynamical parameters are determined in Section \ref{sec:DynamicalParameters}. The clustering procedure is outlined in Section \ref{sec:ClusteringProcedure}. Section \ref{sec:StructureAssociations} explores the clusters and their association to known MW structures. We explore the global properties of the clusters in Section \ref{sec:Global}. Finally, Section \ref{sec:Discussion}
presents a short discussion and perspectives on future directions.

\section{Data}\label{sec:Data}

The B$\&$B survey \citep{Schlaufman2014} forms the basis for the compilation of our data set. The metal-poor candidates were specifically chosen to be bright (\textit{V} $< 14$) stars that exhibit weak molecular mid-IR features around $4.6$ microns compared with near-IR photometry (see \citealt{Schlaufman2014} for further discussion on the candidate selection). Although this method is reasonably successful, it was necessary to obtain validation spectroscopy of the candidates prior to their being used to populate target lists for various ongoing high-resolution spectroscopic studies \citep{Hansen2018,Sakari2018,Ezzeddine2020,Holmbeck2020}.  

A comprehensive explanation of the construction of the spectroscopic sample, along with details of the medium-resolution spectroscopic observations, are discussed in both \citet{Placco2019} and \citet{Limberg2021c}. Here we simply summarize the most pertinent information.  The spectra of the target stars were taken with either the Gemini-North/South (GMOS-N/S) $8.1$-m telescope ($2263$ targets; primarily during bad-weather time), the ESO New Technology (NTT) $3.58$-m telescope ($256$ targets), the Southern Astrophysical Research (SOAR) $4.1$-m telescope ($167$ targets), or the Kitt Peak National Observatory (KPNO) Mayall $4$-m telescope ($73$ targets), over the course of a seven-year campaign from $2014$ to $2020$, resulting in observations for a total of $2760$ bright, metal-poor stellar candidates. The typical wavelength range that these spectra cover is $3200 \lesssim \lambda (\text{\AA}) \lesssim 5800$ for GMOS-N/S targets, $3300 \lesssim \lambda (\text{\AA}) \lesssim 5100$ for NTT targets, $3600 \lesssim \lambda (\text{\AA}) \lesssim 6200$ for SOAR targets, and $3600 \lesssim \lambda (\text{\AA}) \lesssim 6300$ for KPNO targets. The effective resolving power achieved is \textit{R} $\sim 2000$ for GMOS-N/S spectra, \textit{R} $\sim 1200$ for NTT spectra, \textit{R} $\sim 1500$ for SOAR spectra, and \textit{R} $\sim 1800$ for KPNO spectra. For all of the telescopes, the exposure times were chosen such that the Ca II K 3933.3\,{\AA} line has a signal-to-noise ratio (S/N) of at least S/N $\sim 30$ per pixel. These spectra were reduced using standard IRAF \footnote{\url{https://iraf-community.github.io/}} packages \citep{Tody1986,Tody1993}.

The reduced spectra were then used to determine the stellar atmospheric parameters and elemental abundances for the stars using the non-SEGUE Stellar Parameter Pipeline (n-SSPP; \citealt{Beers2014,Beers2017}), which has been adapted to analyze spectra that were not part of the Sloan Extension for Galactic Understanding and Evolution (SEGUE; \citealt{Yanny2009}), originally constructed as the SEGUE Stellar Parameter Pipeline (SSPP; \citealt{Lee2008a,Lee2008b,Lee2011,Lee2013}). The  parameters obtained are the effective temperature (T$_{\rm eff~Spec}$), surface gravity (log \textit{g}), and metallicity ([Fe/H]$_{Spec}$), while the elemental abundances are the carbon abundance ([C/Fe]), and the $\alpha$-element abundance ([$\alpha$/Fe]).  The elemental abundances were obtained by using a $\chi^{2}$ minimization technique for each spectrum and a dense grid of synthetic spectra to choose the optimal set of abundance estimates. The carbon abundance is then adjusted, using the prescription outlined in \citet{Placco2014}, to account for the depletion of carbon along the red giant branch. This corrected carbon abundance ([C/Fe]$_{c}$) is used as the star's natal carbon abundance. The average errors adopted for each of the stellar parameters for the spectra with S/N $\sim 30$ are $\pm 150$ K for $T_{\text{eff}}$; $\pm 0.35$ dex for log \textit{g}, and $\pm 0.20$ dex for [Fe/H], [C/Fe], and [$\alpha$/Fe], with values for each star listed in Table 6 in the Appendix (See \citealt{Lee2008a} for more information on the errors).  Note that the elemental abundances reported here supersede those published in \citet{Placco2019} and \citet{Limberg2021c}\footnote{The [C/Fe] and $\alpha$-element abundances present in \citet{Placco2019} and \citet{Limberg2021c} were found to be systematically offset; the updated values are listed here. The stellar parameters were not affected, in general, although there are occasionally different reported [Fe/H]$_{Spec}$ values reported based on later re-inspections of the spectra by Beers.}.

\begin{figure*}[t]
    \includegraphics[width=\textwidth]{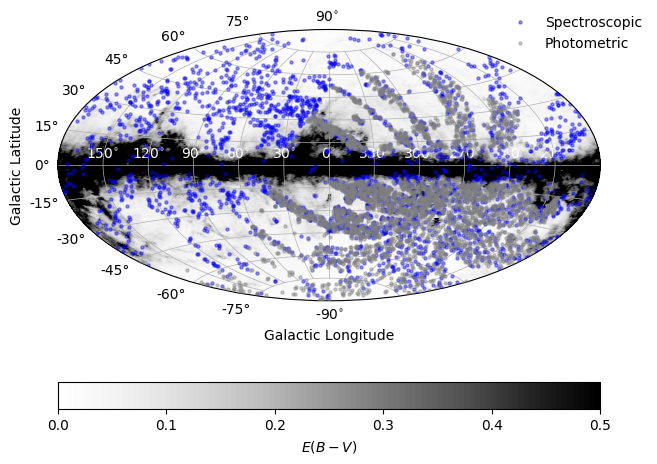}
    \caption{The Galactic positions of the B$\&$B Initial Sample with the spectroscopic subset shown as blue points and the photometric subset as gray points. The Galactic reddening map, taken from \citet{Schlegel1998}, and recalibrated by \citet{Schlafly2011}, is shown in the background on a gray scale with darker regions corresponding to larger reddening.}
    \label{fig:galactic_map}
\end{figure*}

The rest of our sample is assembled from recent photometric estimates of temperature, luminosity classes, and metallicity for candidate stars from the B$\&$B survey based on the procedure described by \citet{Huang2021c}. This study made use of re-calibrated zero-points \citep{Huang2021b} in the narrow- and medium-band photometry obtained by the Sky Mapper Southern Survey (SMSS; \citealt{Wolf2018}) Data Release 2 (DR2; \citealt{Onken2019}) along with broad-band photometry from Gaia EDR3 \citep{GaiaCollaboration2021}. The average errors adopted for each of the stellar parameters for the photometric portion of the sample are $\pm 64$ K for $T_{\text{eff}}$, and $\pm 0.20$ dex for [Fe/H], with values for each star listed in Table 6 in the Appendix (See \citealt{Huang2021c} for more information on the errors). A total of $4935$ stars have available photometric estimates of effective temperature and metallicity; $1130$ of these stars also have available
medium-resolution spectra. \citet{Huang2021c} have compared the photometric estimates of $T_{\text{eff}}$ and [Fe/H] from their catalog with several medium-resolution studies (including stars from the Pristine Survey follow-up reported by \cite{Aguado2019}, and from the Best \& Brightest sample reported by \citealt{Limberg2021c}) and high-resolution studies (near-IR spectroscopy from APOGEE DR14 \citep{Abolfathi2018} and DR16 \citep{Ahumada2020}, and optical spectroscopy from a number of individual papers), and find generally excellent agreement.
The spatial distribution of these sub-samples of B$\&$B stars in Galactic coordinates can be seen in Figure~\ref{fig:galactic_map}. Hereafter, we refer to the full sample of stars as the Initial Sample.  Trimming of this sample to obtain a subset of stars suitable for our dynamical analysis is described below. 

\begin{figure*}[t]
    \includegraphics[width=\textwidth]{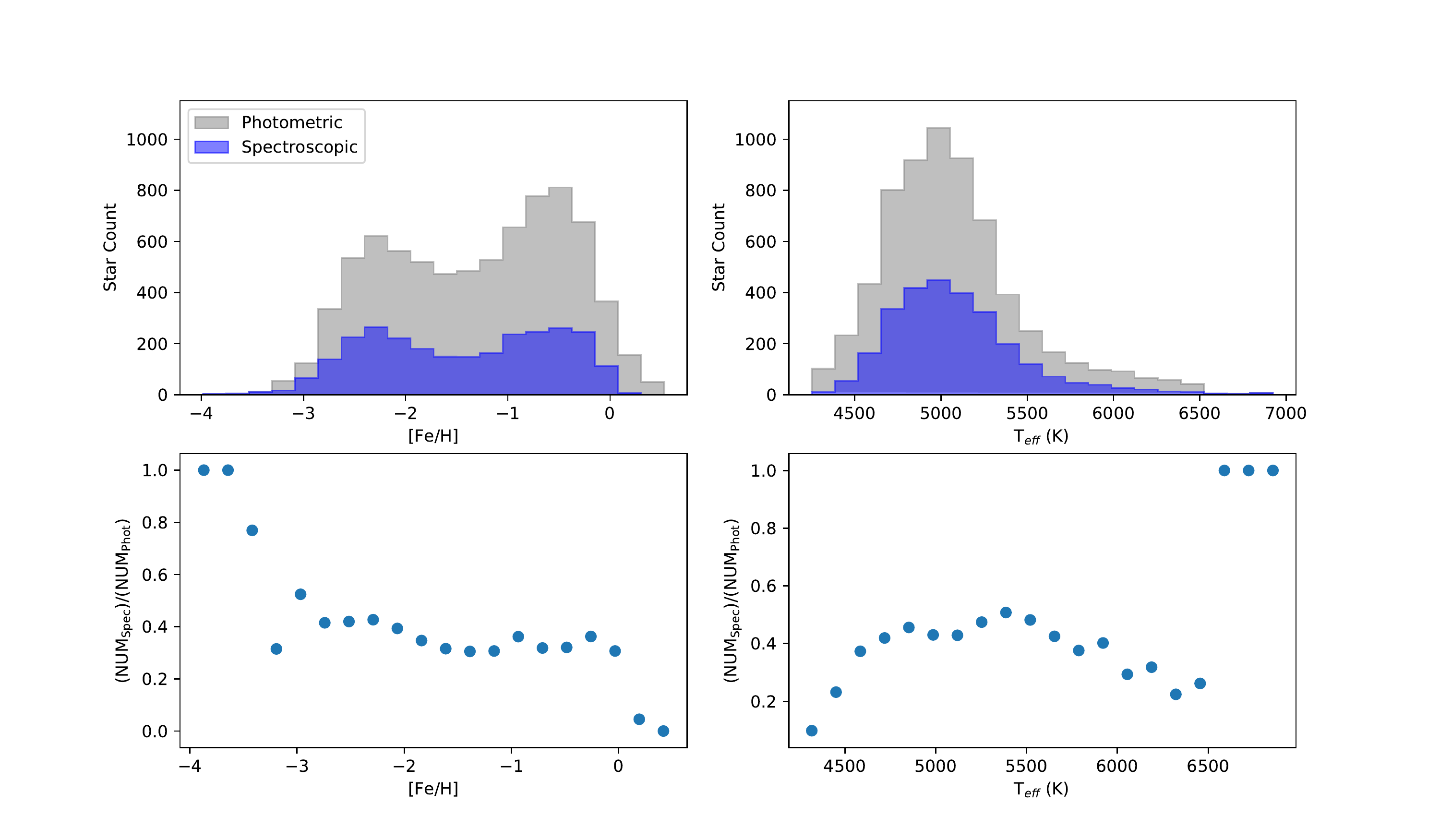}
    \caption{Top Left Panel: Histogram of the metallicity for the spectroscopic ([Fe/H]$_{Spec}$, gray) and photometric ([Fe/H]$_{Phot}$, blue) stars for the Initial Sample. Top Right Panel: Histogram of the effective temperatures for the spectroscopic (T$_{\rm eff~Spec}$, gray) and photometric (T$_{\rm eff~Phot}$, blue) stars for the Initial Sample. Bottom Panels: The relative number of stars in the spectroscopic sample to the number in the photometric sample in each bin corresponding to the histogram above. The edges are sparsely populated, leading to the extreme variation shown. The middle region of these panels show that the relative number of spectroscopic to photometric sample stars are similar across the range of the parameter space used in our dynamical analysis.}
    \label{fig:metallicity_teff_hist}
\end{figure*}

Figure~\ref{fig:metallicity_teff_hist} shows the distributions of [Fe/H] and $T_{\rm eff}$ estimates obtained from the spectroscopic and photometric sub-samples in the top panels, respectively.  As can be appreciated from inspection of this figure, these subsets are (with the exception of their slightly different adopted cutoffs for metallicity and temperature) essentially identical. The bottom panels show the number of stars in the spectroscopic sample relative to the number of stars in the photometric sample in each of the bins of the histograms in the top panels. The deviations at the edges of the [Fe/H] distribution are primarily a result of the small number statistics for each sample, with the low-metallicity end being dominated by spectroscopic measurements, and the high-metallicity end being dominated by photometric measurements. The opposite is true for the effective temperature; the low-temperature range is dominated by the photometric sample, and the high-temperature range of the photometric sample does not extend beyond $6500$\,K. The middle region of the bottom panels show that the relative number of spectroscopic to photometric stars is similar across the range of the parameter space used in our dynamical analysis described below.

\begin{figure}[t]
    \includegraphics[width=0.48\textwidth]{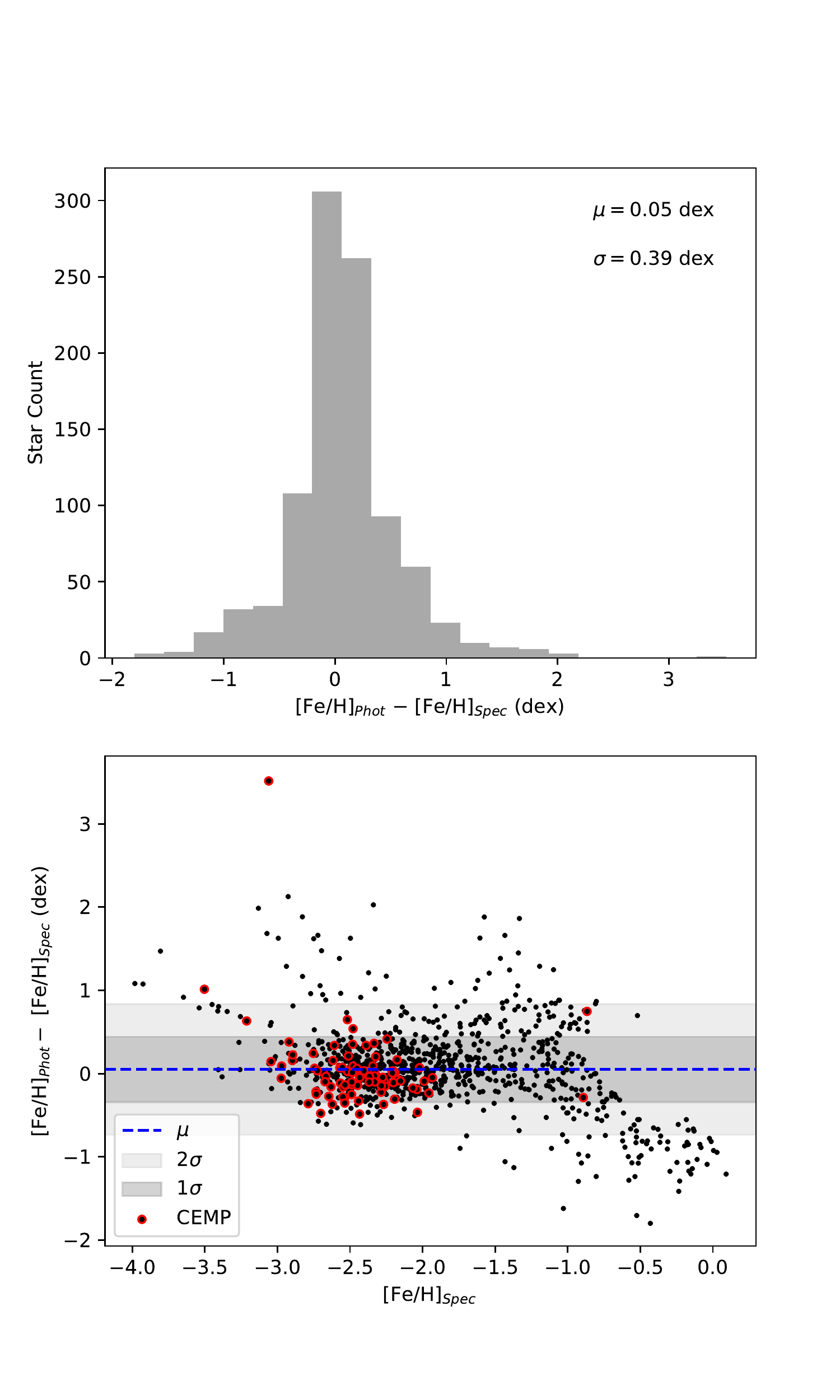}
    \caption{Top Panel: Histogram of the residuals of the difference between the photometric metallicity values ([Fe/H]$_{Phot}$) and the metallicity values obtained from medium-resolution spectra (Fe/H]$_{Spec}$) for the Initial Sample. The biweight locations and scales are noted. Bottom panel: The residuals between [Fe/H]$_{Phot}$ and [Fe/H]$_{Spec}$, as a function of [Fe/H]$_{Spec}$, for the Initial Sample. CEMP stars are indicated with red outlined points. The blue dashed line is the biweight location, while the shaded regions represent the first (1$\sigma$) and second (2$\sigma$) biweight scale ranges.  The apparent systematic discrepancy at high metallicity in this panel is likely due to difficulties in the n-SSPP estimates in this range (see text).}
    \label{fig:metallicity_diff_hist}
\end{figure}

The biweight estimators of location and scale (see \citealt{Beers1990}) for the metallicity residuals determined from the medium-resolution spectra and the photometric metallicity yield $\mu = 0.05$ dex and $\sigma = 0.39$ dex; a comparison between the two subsets is shown in Figure~\ref{fig:metallicity_diff_hist}.  As noted by \citet{Huang2021c}, stars that have enhanced carbon may have estimated photometric metallicities that are somewhat higher than the spectroscopic determinations, due to molecular carbon features affecting the blue narrow/medium-band filters $v$ and $u$ from SMSS (particularly for cooler carbon-enhanced stars).  Stars with [C/Fe]$_c > +0.7$ are indicated with red circles around the dots shown in the bottom panel of Figure~\ref{fig:metallicity_diff_hist}. We note that, when these stars are removed from the sample, similar residuals and offsets are found as for the entire sample.
From inspection of this figure, there appears to be a systematic discrepancy in the metallicity estimates in the metal-rich region.  We attribute this to difficulties encountered by the n-SSPP estimates\footnote{The n-SSPP uses a subset of the estimators employed by the SSPP, which has been demonstrated previously to work very well in this high-metallicity range.}, rather than the photometric estimates, which have been shown by \citet{Huang2021c} to have excellent performance in this metallicity regime.  As noted below, we only retain stars with [Fe/H] $\leq -0.8$ in the Final Sample, so these stars will not affect our subsequent analysis.  

\begin{figure}[t]
    \includegraphics[width=0.48\textwidth]{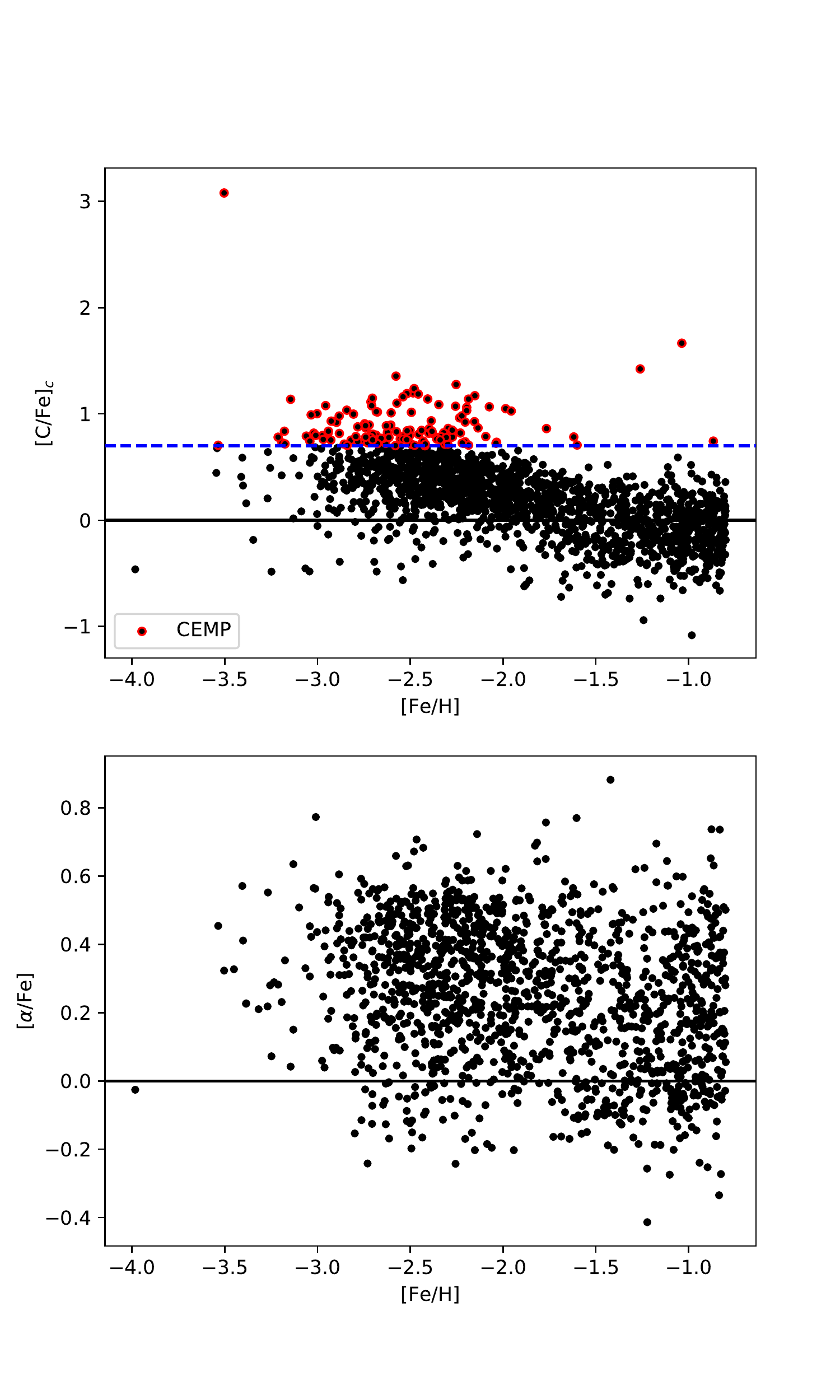}
    \caption{Top Panel: The corrected carbon abundance ([C/Fe]$_{c}$) of the Initial Sample, as a function of the final metallicity ([Fe/H]). The CEMP cutoff ([C/Fe]$_{c} = +0.7$) is noted as the blue dashed line; CEMP stars are indicated with red outlined points. The Solar value is indicated as the solid black line. Bottom panel: The $\alpha$-element abundance ([$\alpha$/Fe]) of the Initial Sample, as a function of the final metallicity ([Fe/H]). The Solar value is indicated with a solid black line.}
    \label{fig:abundances}
\end{figure}

\begin{figure}[t]
    \includegraphics[width=0.48\textwidth]{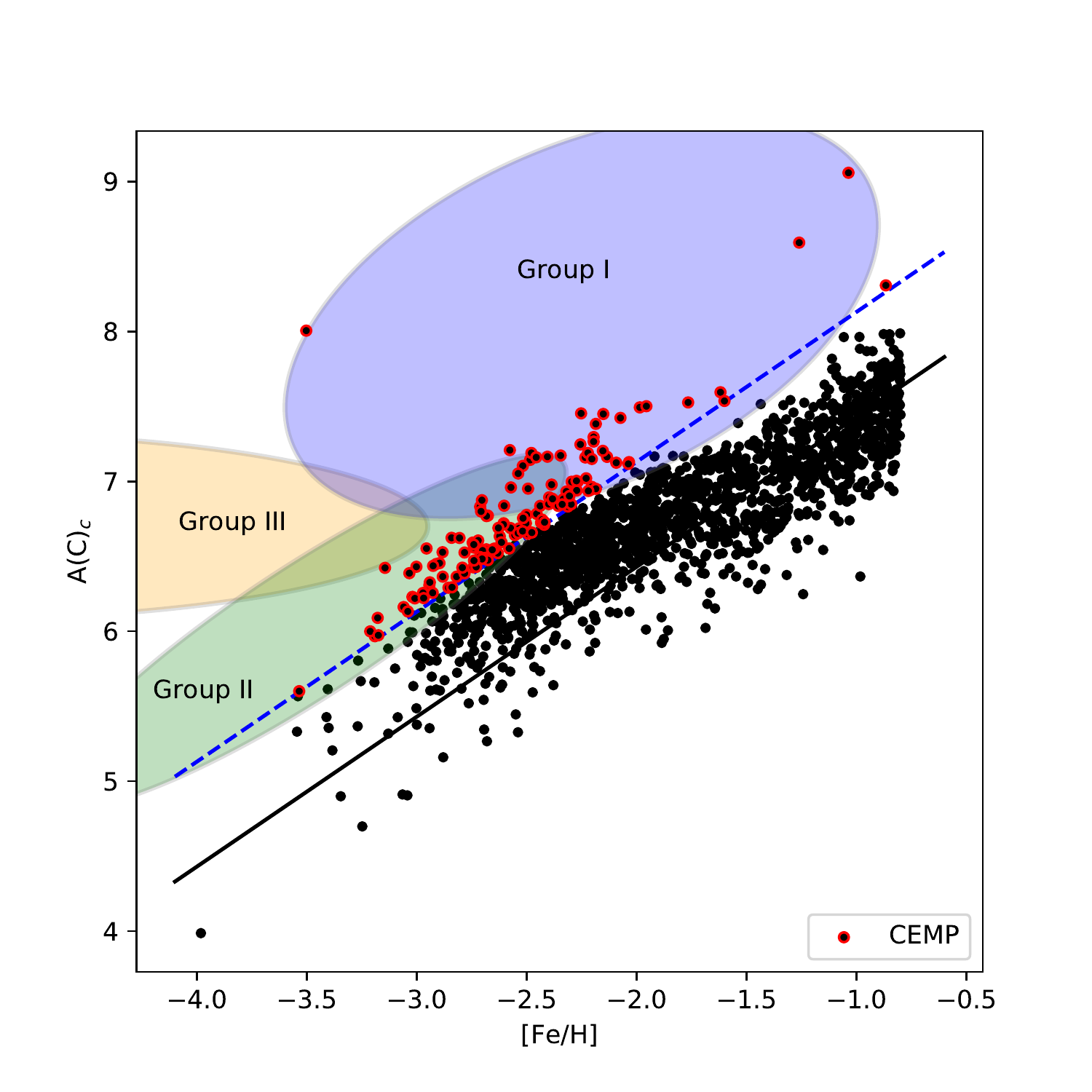}
    \caption{The Yoon-Beers Diagram of $A{\rm (C)}_c$, as a function of [Fe/H], for stars in the Initial Sample with [Fe/H] $\leq -0.8$. The CEMP cutoff ([C/Fe]$_{c} = +0.7$) is indicated with a blue dashed line. The CEMP stars ([C/Fe]$_{c} > +0.7$) are shown as red outlined points. [C/Fe] $= 0$ is indicated with a solid black line. The ellipses represent the three different morphological groups of CEMP stars (See Figure~1 in \citealt{Yoon2016} for a comparison and more information).}
    \label{fig:yoon_beers}
\end{figure}

The [C/Fe]$_{c}$ and [$\alpha$/Fe] estimates for stars in the Initial Sample are shown in Figure~\ref{fig:abundances}. The Yoon-Beers diagram of A$(C)_{c}$ vs. [Fe/H] for Initial Sample stars with [Fe/H] $\leq -0.8$ is shown in Figure~\ref{fig:yoon_beers}.

For stars that have both spectroscopic and photometric stellar parameters, we perform a procedure to obtain the parameters available to produce final adopted estimates of [Fe/H] and $T_{\rm eff}$. If the absolute difference in metallicity between the two samples ([Fe/H]$_{Spec}$ $-$ [Fe/H]$_{Phot}$) is less than $0.5$ dex, then we average the two sets of parameters. If not, then a choice is made to adopt either the spectroscopic or photometric parameters based on visual inspection of the spectrum for a given star. For the stars with available spectroscopy, we then adjust estimates of the carbon and $\alpha$-element abundances from the n-SSPP based on the final adopted [Fe/H].

\begin{figure*}
    \includegraphics[width=\textwidth]{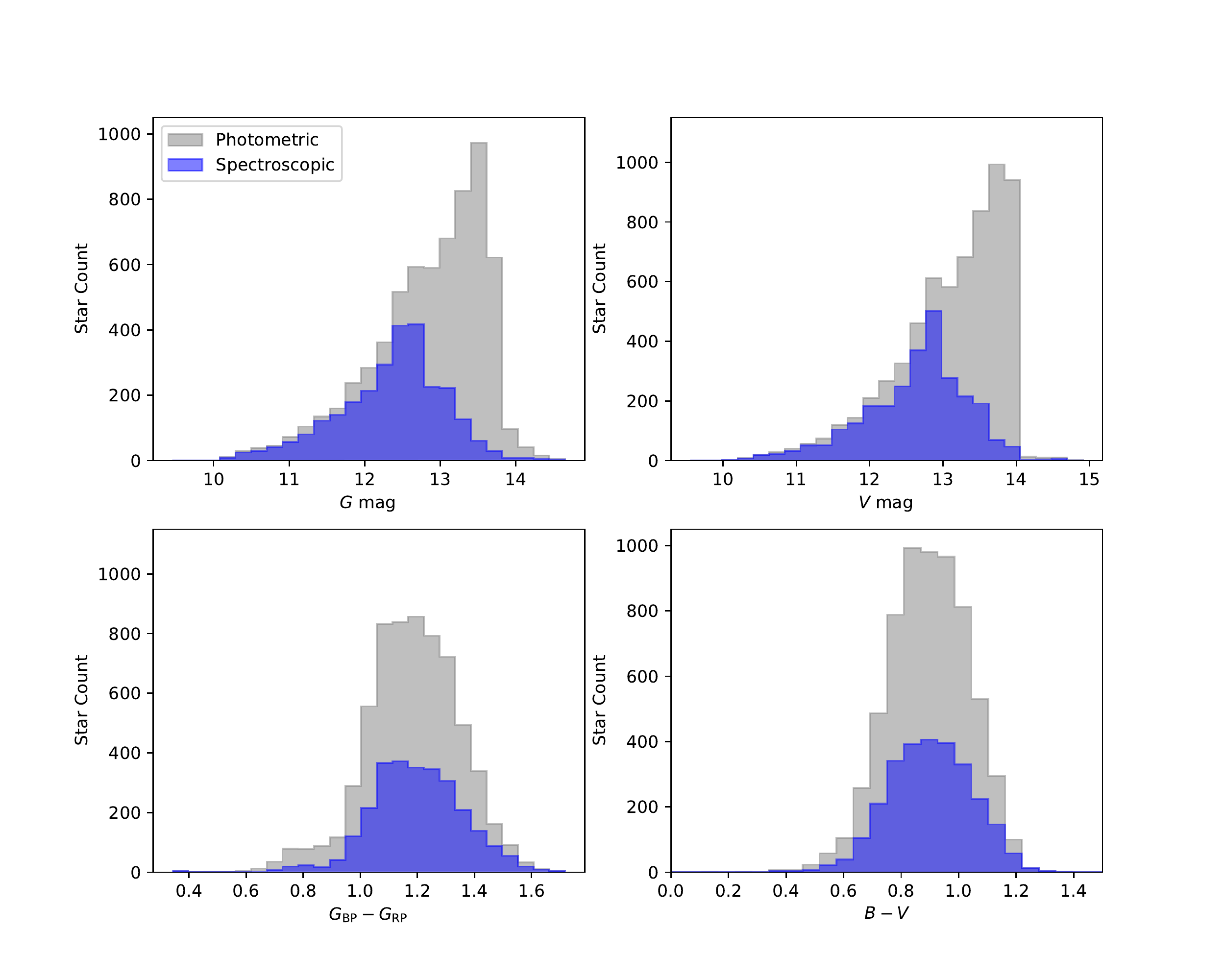}
    \caption{Top Left Panel: Histogram of the $G_{\text{mag}}$ for the Initial Sample. Top Right Panel: Histogram of the $V_{\text{mag}}$ for the Initial Sample. Bottom Left Panel: Histogram of the $G_{\text{BP}} - G_{\text{RP}}$ color for the Initial Sample. Bottom Right Panel: Histogram of the $B - V$ color for the Initial Sample; we note that there are a few stars not included in the range shown. All Panels: The spectroscopic subset of the Initial Sample is represented as a blue histogram; the photometric subset is represented as a gray histogram.}
    \label{fig:mag_hist}
\end{figure*}

The stars were then cross-matched with Gaia Early Data Release 3 (EDR3; \citealt{GaiaCollaboration2021}) using a $5 \arcsec$ radius to find their $6$-D astrometric parameters. The astrometry for B$\&$B \citep{Schlaufman2014} comes from APASS DR6 \citep{Henden2012}, 2MASS \citep{Skrutskie2006}, and AllWISE \citep{Wright2010,Mainzer2011}. The astrometry of these surveys is usually good to $5 \arcsec$ or better, so we use a $5 \arcsec$ radius to capture the matches. To validate the match for each star, confirmation was performed by checking that the stellar magnitudes agreed to within $0.5$ mag between the sources. The initial sample was mostly taken from \textit{V} magnitudes supplied by the AAVSO Photometric All Sky Survey (APASS; \citealt{Henden2014}) Data Release 9 (DR9; \citealt{Henden2016}), with various other sources listed in the Appendix tables supplying the rest. The corresponding matches were then compared with the \textit{V} magnitude utilizing the transformation from Gaia magnitudes \textit{G}, \textit{G}$_{\text{BP}}$, and \textit{G}$_{\text{RP}}$ in EDR3 provided in Table C.2 of \citet{Riello2021}.  A comparison of the magnitudes can be seen in Figure \ref{fig:mag_hist}. From inspection of the figure, although the color ranges of the two subsets are essentially identical, the magnitudes for the two subsets differ in the sense that the photometric sub-sample includes stars that are fainter than those from the spectroscopic sub-sample.

\begin{figure}
    \includegraphics[width=0.5\textwidth]{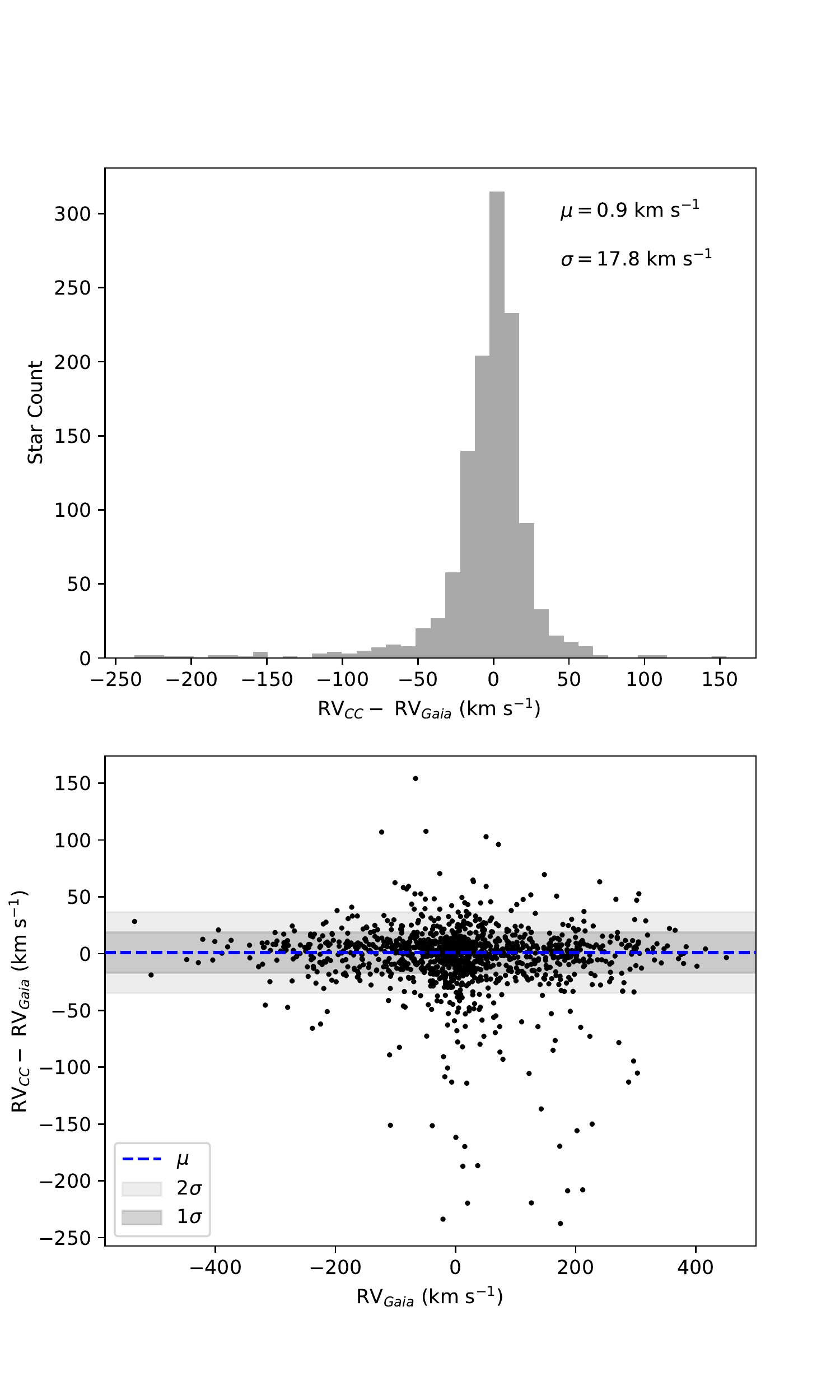}
    \caption{Top Panel: Histogram of the residuals of the difference between the CC radial velocities and the Gaia EDR3 values in the Initial Sample. The biweight location and scale are noted. Bottom panel: The residuals between the CC and Gaia EDR3 radial velocities, as a function of the Gaia radial velocities. The blue dashed line is the biweight location of the residual difference ($\mu = 0.9$ km s$^{-1}$), while the shaded regions represent the first (1$\sigma$ = 17.8 km s$^{-1}$), and second (2$\sigma$ = 35.6 km s$^{-1}$) biweight scale ranges.}
    \label{fig:rv_hist}
\end{figure}

Radial velocities, parallaxes, and proper motions for each star are taken from Gaia EDR3, when available. Note that the radial velocities for the stars in Gaia EDR3 are available for only about half ($48\%$) of the Initial Sample. For the subset of stars with available medium-resolution spectra, each of the spectra was cross-correlated (CC) with a radial-velocity standard star, and corrected to the rest frame. Typical errors for Gaia EDR3 radial velocities are ${\sim} 1$ km s$^{-1}$; The CC radial velocities for the medium-resolution spectra have ${\sim} 5$ km s$^{-1}$ errors. The top panel of Figure~\ref{fig:rv_hist} shows a histogram of the residual differences between the CC radial velocities and the Gaia EDR3 values. The biweight location and scale of these differences are $\mu = 0.9$ km s$^{-1}$ and $\sigma = 17.8$ km s$^{-1}$, respectively. The bottom panel of this figure shows the residuals between the CC and Gaia EDR3 radial velocities, as a function of the Gaia radial velocities. The blue dashed line is the biweight location, while the shaded regions represent the first (1$\sigma$) and second (2$\sigma$) biweight scale ranges.  We expect that many of the stars with residuals outside the 2$\sigma$ range are binaries, or the wavelength solution of the spectrum is off, causing an improper estimation of stellar parameters and radial velocities.

\begin{figure}
    \includegraphics[width=0.5\textwidth]{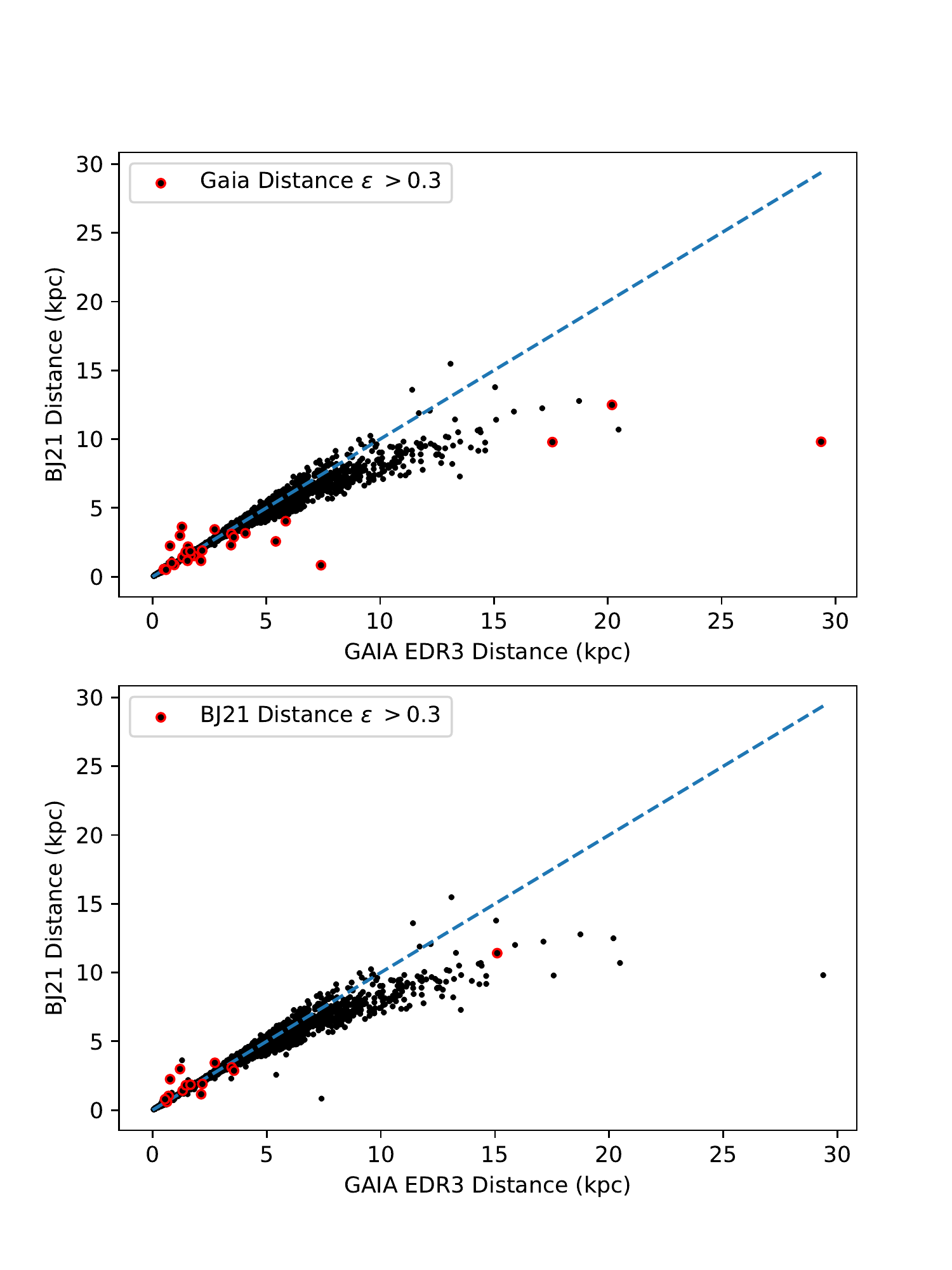}
    \caption{Top Panel: Comparison between the Gaia EDR3 corrected inverse-parallax distance and BJ21 distances in the Initial Sample. Stars with the red outlined points indicate that the relative distance error in Gaia EDR3 is $\epsilon > 0.3$. Bottom panel: The same comparison as the top panel, but with the red outlined points indicating the relative distance error in BJ21 is $\epsilon > 0.3$. In both plots the dashed line indicates a one-to-one comparison between the two samples.}
    \label{fig:dist_comp}
\end{figure}

The distances to the stars are determined either through the (corrected) inverse parallax given by Gaia EDR3 or the Bailer-Jones distance estimate (BJ21; \citealt{Bailer-Jones2021}). The Gaia EDR3 distance is given by the inverse of the parallax, after first being corrected for the $0.026$ mas offset noted by \citep{Huang2021a}. Parallax values in our Initial sample from EDR3 have an average error of around $0.04$ mas.  The BJ21 distances are determined by a Bayesian approach utilizing the EDR3 parallax, magintude, and color \citep{Bailer-Jones2021}. The errors are presented for each star in the tables provided in the Appendix.
We prioritize the BJ21 distances when the relative error (the error divided by the reported value), $\epsilon$, is $ \epsilon < 30\%$. 
If the BJ21 relative error is $\geq 30\%$, then we adopt the inverse parallax distance if the relative error is $ \epsilon < 30\%$. If only one distance estimator is available, then we adopt it. If both distances, or the only available distance, have $\epsilon \geq 30\%$, then we discard the star from the dynamical analysis below. Note that in Figure~\ref{fig:dist_comp} the inverse-parallax distances obtained from Gaia EDR3 tend to overestimate the distances compared to the BJ21 approach, especially when the distance is greater than $5$ kpc. The proper motions in our Initial sample from Gaia EDR3 have an average error of $39$ $\mu$as yr$^{-1}$.

\section{Dynamical Parameters}\label{sec:DynamicalParameters}

The orbital characteristics of the stars are determined using the Action-based GAlaxy Modelling Architecture
\footnote{\url{http://github.com/GalacticDynamics-Oxford/Agama}} (\AGAMA) package \citep{Vasiliev2019}. The orbits are determined with the Solar position of ($-8.249$, $0.0$, $0.0$) kpc \citep{GravityCollaboration2020} and Solar peculiar motion in (U,W) as ($11.1$,$7.25$) km s$^{-1}$ \citep{Schonrich2010} with Galoctocentric Solar azimuthal velocity  \textit{V} $= 250.70$ km s$^{-1}$ determined from \citet{Reid2020}. Note that these values do not depend on the MW gravitational potential, specifically the V-component of the Solar motion, which normally relies on the Local Standard of Rest (LSR) whose value changes with differing choices of MW gravitational potentials. To calculate the orbits we require a MW gravitational potential, chosen to be the \MWMMXVII ~potential \citep{McMillan2017} for consistency with previous work \citep{Gudin2021,Limberg2021a}. This potential models the MW with six axisymmetric components: bulge, dark-matter halo, thin- and thick- stellar disks, an HI gas disk, and a molecular gas disk.

The astrometric parameters needed to determine the dynamical parameters from \AGAMA ~are the positions (RA and DEC), proper motions in the position directions (PM$_{\text{RA}}$ and PM$_{\text{DEC}}$), the errors for both PM$_{\text{RA}}$ and PM$_{\text{DEC}}$ (along with the correlation coefficient, PM$_{\text{corr}}$), the radial velocity and corresponding error, and the distance and corresponding error. The procedure for obtaining these values are described in Section \ref{sec:Data} above.

These astrometric parameters are run through the orbital integration process in \AGAMA. They undergo static orbits for $100$ Gyr with the stellar orbital positions and velocities being sampled at a rate of every $1$ Myr. The long integration time is adopted because of the presence of stars on orbits that can reach far out into the stellar halo;  this is done in order to capture accurate quantities of pericentric (closest) and apocentric (farthest) distances (\textit{r}$_{\peri}$ and \textit{r}$_{\apo}$, respectively), along with the maximum height above the Galactic plane (Z$_{\maxtext}$).
The \AGAMA ~procedure then calculates the orbital energy, cylindrical positions and velocities (v$_{\text{r}}$,v$_{\phi}$,v$_{\text{z}}$), angular momentum, and cylindrical actions using the \MWMMXVII ~potential. 
The eccentricity of the orbit is determined by $ecc = (r_{\apo} - r_{\peri})/(r_{\apo} + r_{\peri})$.

The orbital energy of a star, E, is the sum of the kinematic energy and the gravitational potential energy. The kinematic energy is determined from the total velocity of the star, while the gravitational potential energy is determined through the adopted gravitational potential of the MW, in this case \MWMMXVII, and the stellar position. A star is considered bound to the Galactic potential if the orbital energy is less than zero, while an unbound star has an orbital energy greater than zero.

\citet{Binney2012} introduced an efficient algorithm to compute action in an axisymmetric potential. The actions (J$_{\text{r}}$,J$_{\phi}$,J$_{\text{z}}$) describe the extent of the stellar orbit, and are summarized as follows. The radial action, J$_{\text{r}}$, indicates the star's overall movement in the radial dimension over the course of its orbit, and can be J$_{\rm r} \geq 0$, with J$_{\rm r} = 0$ indicating a circular orbit. For a given orbital energy, the higher the stellar orbit's eccentricity, the higher the radial action. The azimuthal action, J$_{\phi}$, gauges the guiding center radius or the mean orbital radius of the stellar orbit, and can be any value, either positive or negative. For an axisymmetric potential, which is the case for \MWMMXVII, the azimuthal action is given by J$_{\phi} = -$L$_{\text{z}}$ for a right-handed coordinate system based around the MW's rotation. 
The negative sign is introduced so that prograde stars have positive J$_{\phi}$ and retrograde stars have a negative J$_{\phi}$. 
The vertical action, J$_{\text{z}}$, is the change in the star's height above the Galactic plane over the course of the stellar orbit. The vertical action is given by $J_{\rm z} \geq 0$, with J$_{\rm z} = 0$ being a planar orbit. 
For a given orbital energy, the larger the maximum distance from the Galactic plane a star travels, the larger the vertical action. The quantities used throughout are the specific orbital energy and actions. These are the orbital energy and actions divided by the stellar mass (though we simply refer to them as the orbital energy and actions). For reference, we note the Solar orbital energy is $-1.51 \times 10^{5}$ km$^{2}$ s$^{-2}$, and its actions are ($10.1$, $2.07 \times 10^{3}$, $0.345$) kpc km s$^{-1}$ in our adopted Galactic model.

This procedure obtains the orbital parameters if the astrometric parameters are precisely described by the given values. However, these values have errors associated with them, so a method must be developed to estimate the errors in the orbital parameters. This is accomplished through a Monte Carlo sampling over the errors in the astrometric parameters. The proper motion errors, described above, are employed to create a covariance matrix, while the errors for the radial velocity and distance values are used as is. The errors are assumed normally distributed, and each Monte Carlo sample draws a random value for each parameter based on the mean (the adopted value) and the standard deviation for the the error. Once each value is drawn, the procedure to perform the orbital integration process is executed and outputs are stored for each sample. This is done $1000$ times in order to robustly evaluate the statistical uncertainties associated with the orbital properties. Once all the orbital values are collected, the mean and standard deviation of the $1000$ trials are determined, and reported for the orbital parameters. 

An inspection is performed to identify stars that are not suitable for the following dynamical analysis. The Initial Sample of $6445$ stars contains stars that either not metal-poor, or have poorly determined stellar parameters. A cut was performed on the final adopted metallicities, only retaining stars with [Fe/H] $\leq -0.8$. This results in $4068$ stars. Stars that are unbound from the MW (E$ > 0$), along with stars that do not have the full $6$-D astrometric parameters of position, radial velocity, distance, and proper motions were then cut. A total of $2026$ stars lacked radial velocities, while $12$ of these stars lacked any distance estimates. All $25$ stars that did not have proper motions also did not have any distance estimates. Finally, in order to obtain accurate orbital dynamics, we cut the $213$ stars with differences in their CC radial velocities compared to the Gaia radial velocities that lie outside the 2$\sigma$ region shown in Figure~\ref{fig:rv_hist}. Most of these stars are expected to be binaries, although some could also have spectra with poor wavelength solutions (which would also compromise stellar-parameter estimates). Application of this cut leaves a total sample of $2526$ stars to perform the following analysis. The dynamical parameters of the stars with orbits determined are listed in Table 7 in the Appendix; we refer to this as the Final Sample.

\begin{figure}[t]
    \includegraphics[width=0.49\textwidth]{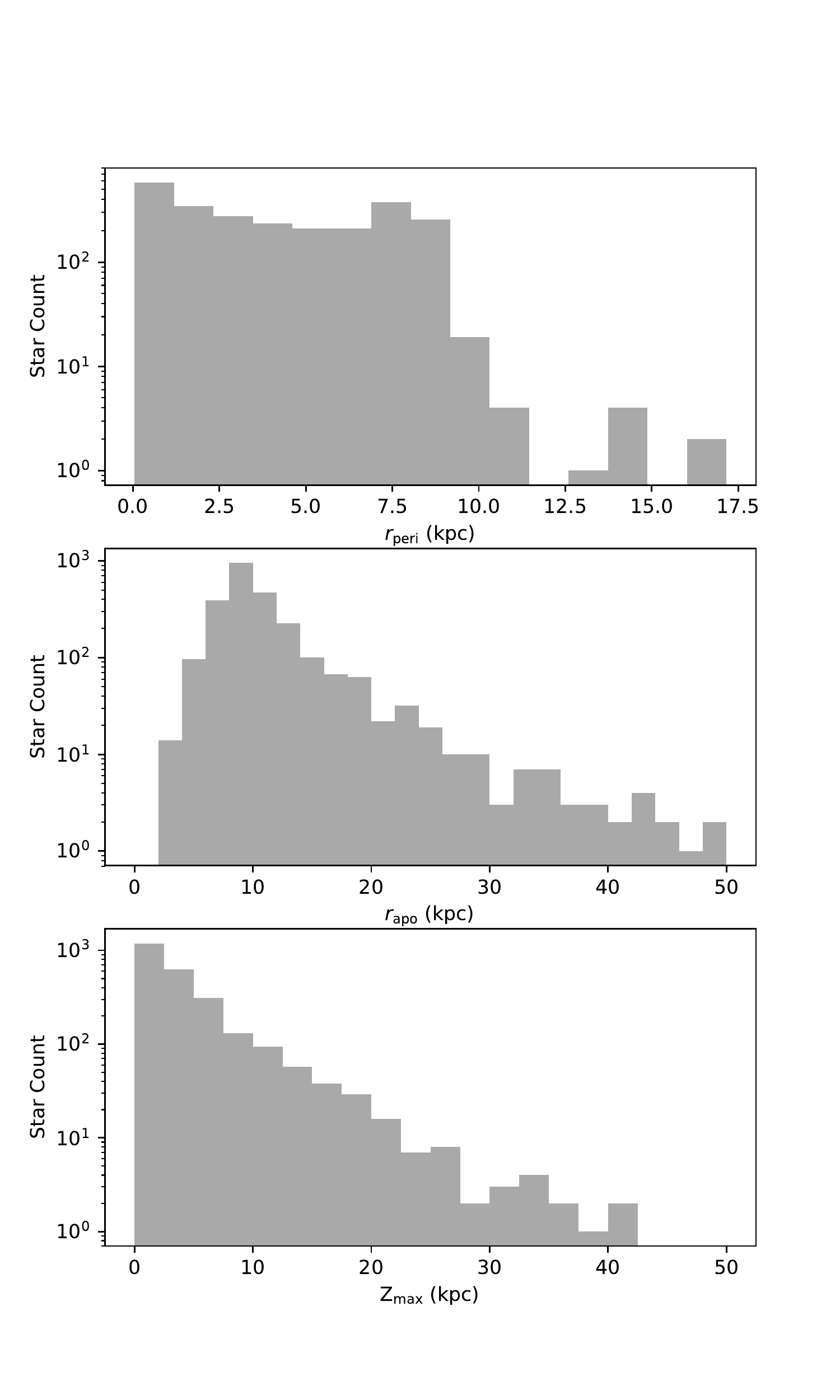}
    \caption{Logarithmic histograms of the orbital parameters \textit{r}$_{\peri}$ (top), \textit{r}$_{\apo}$ (middle), and Z$_{\maxtext}$ (bottom) for the B$\&$B Final Sample. Note that a few stars have \textit{r}$_{\apo}$ and Z$_{\maxtext}$ outside the range shown in the panels.}
    \label{fig:orb_dist}
\end{figure}

Figure~\ref{fig:orb_dist} provides histograms of \textit{r}$_{\apo}$ (top), \textit{r}$_{\peri}$ (middle), and Z$_{\maxtext}$ (bottom) for the Final Sample.  From inspection of this figure, it is clear that the majority of the stars in this sample occupy orbits that take them inside the inner-halo region, but they also explore regions well into the outer-halo region, up to $\sim 50$ kpc away.

\bigskip
\bigskip
\section{Clustering Procedure}\label{sec:ClusteringProcedure}

\citet{Helmi2000} were among the first to suggest the use of integrals of motion, in their case orbital energy and angular momenta, to find substructure in the MW using the precision measurements of next-generation surveys that were planned at the time. \citet{McMillan2008} suggested the use of actions as a complement to the previously suggested orbital energy and angular momenta, with only the vertical angular momentum being invariant in an axisymmetric potential. Previously, actions were difficult to determine outside of a spherical potential, but recent progress allowed actions to be obtained in realistic MW potentials \citep{Binney2012}. Most recently, many authors have employed the orbital energy and cylindrical actions (E,J$_{r}$,J$_{\phi}$,J$_{z}$) to determine the substructure of the MW using both Gaia DR1 and DR2 measurements \citep{Helmi2017,Myeong2018b,Myeong2018c,Roederer2018,Naidu2020,Yuan2020a,Yuan2020b,Gudin2021,Limberg2021a}.

The unsupervised clustering algorithm \HDBSCAN ~\citep{Campello2013} has been employed by several authors to determine substructure in the form of dynamical clusters \citep{Koppelman2019a,Gudin2021,Limberg2021a}. We use \HDBSCAN ~as well, in order to perform a cluster analysis over the orbital energy and cylindrical actions from the Final Sample obtained through the procedure outlined in Section \ref{sec:DynamicalParameters}. The \HDBSCAN ~algorithm operates through a series of calculations that are able to separate the background noise from denser clumps of data in the dynamical parameters. Below we describe the main aspects of the 
\HDBSCAN~algorithm\footnote{For a detailed description of the \HDBSCAN ~algorithm visit: \url{https://hdbscan.readthedocs.io/en/latest/how_hdbscan_works.html}}.

\subsection{The \HDBSCAN~Algorithm}\label{sec:HDBSCAN}

The identification of dynamical clusters requires the construction of a set of cores, each of which have a metric applied to determine their relationship to one another. This metric forms the basis of a minimal spanning tree, which when condensed will allow clusters of the dense cores to fall out of the algorithm; everything else is relegated to background noise. One clear advantage of using \HDBSCAN ~is that the algorithm makes no underlying model assumptions about the data, while allowing the user to control the process using physically significant parameters, as described below.

For our run of \HDBSCAN, ~we utilize the following parameters: \verb min_cluster_size ~$= 5$, \verb min_samples ~$= 5$, \verb cluster_selection_method ~$=$ \verb 'leaf' , and \verb prediction_data ~$=$ \verb True . The \verb min_cluster_size ~parameter determines the minimum number of stars that will be considered to form a cluster. The \verb min_samples ~parameter regulates how conservative the \HDBSCAN ~algorithm produces clusters. Larger values of the \verb min_samples ~parameter are more conservative, and will assign more data to noise compared to lower values. The \verb min_samples ~parameter is defaulted to the value of the \verb min_cluster_size ~parameter. \citet{Limberg2021a} explored the clusters when changing the \verb min_samples ~parameter to various values, and determined the default is the best option based on how small the stellar clusters already are for this brand of clustering (around 30 stars in the largest cluster). Setting the \verb cluster_selection_method ~parameter to \verb 'leaf' ~allows the clusters to select smaller, homogeneous associations compared to the Excess of Mass, or \verb 'eom' , setting which permits larger, more diffuse associations to fall out of the clustering process. Since the goal of this particular clustering procedure is to find stars that have strongly similar orbital characteristics, the \verb 'leaf' ~setting is preferred. The \verb prediction_data ~parameter is set to \verb True, and is explained in more detail in the following paragraph. With these parameters set, the final step is to transform the data in a way that \HDBSCAN ~can perform the algorithm. The biweight normalization \citep{Beers1990} is applied to each of the orbital energy and cylindrical actions sets. Using the biweight normalization allows for both a more robust and resistant measure of location and scale against outliers, compared to determining the normal distribution mean and standard deviation. The stars are then processed through the \HDBSCAN ~algorithm with the above parameters and scaled data sets. Once complete, each star is characterized through either the cluster number they are assigned to, or $-1$, which represents stars assigned to the background noise. 

As is the case for the orbital parameters, the dynamical clusters require a procedure to estimate the associated errors. We again take $1000$ Monte Carlo samples of the orbital energy and actions using a normal distribution, and for each draw we redetermine which cluster a star belongs to. Usually, if we use a new \HDBSCAN ~object, then the clusters formed are entirely different compared to the original clustering output produced by the nominal values of the orbital energy and actions. In order for the \HDBSCAN ~object to remember the parameter space of the original clusters, we set the \verb prediction_data ~parameter to \verb True . Then, when \HDBSCAN ~is performed on each Monte Carlo sample, the new values of orbital energy and actions are compared to the parameter space already determined by the nominal \HDBSCAN ~clustering procedure, and subsequently put into the clusters with which they best correspond. We then gather the resulting cluster outputs for each star and choose the cluster with which the star is best associated. This is done by extracting the assigned cluster mode from the $1000$ trials for each star. Once the predominant cluster is found, we determine the probability of the star actually being in the cluster by dividing the number of times the star fell into the cluster out of the $1000$ Monte Carlo trials. If the star survives the minimum probability of $20\%$ (consistent with the choice made by \citealt{Gudin2021} and \citealt{Limberg2021a}), then we assign it to the cluster. Since clusters can drop below the minimum cluster size using this procedure, only clusters that have at least the minimum cluster size are kept through the rest of the analysis, while the dropped clusters' stars are assigned to noise.  

Table~\ref{tab:cluster_summary} provides a listing of the Dynamically Tagged Groups (DTGs) identified by this procedure, along with their numbers of member stars, confidence values, and associations described below.  Note that, although a minimum confidence value of 20\% was employed, the actual minimum value found for these DTGs is 36.5\%.  The DTGs and CDTGs are identified using the nomenclature introduced by \citet{Yuan2020b}, to which we refer the interested reader. 

Table~\ref{tab:cluster_results_stub} lists the stellar members of the identified DTGs, along with their final values of [Fe/H], [C/Fe], [C/Fe]$_c$, and [$\alpha$/Fe], where available.  The last line in the listing for each DTG gives the mean and dispersion (both using biweight estimates) for each quantity.

Table~\ref{tab:cluster_orbit_stub} lists the derived dynamical parameters derived by \AGAMA ~used in our analysis.

\section{Structure Associations}\label{sec:StructureAssociations}

Associations between the newly identified DTGs are now sought between known MW structures, including large-scale substructures, previously identified dynamical groups, stellar associations, globular clusters, and dwarf galaxies.

\input{Tables/cluster_summary_table}

\input{Tables/cluster_stellar_results_stub_table}

\input{Tables/cluster_orbital_stub_table}

\subsection{Milky Way Substructures}\label{subsec:MWSubstructure}

\begin{figure*}[t]
    \centering
    \includegraphics[width=0.98\textwidth,height=0.98\textheight,keepaspectratio]{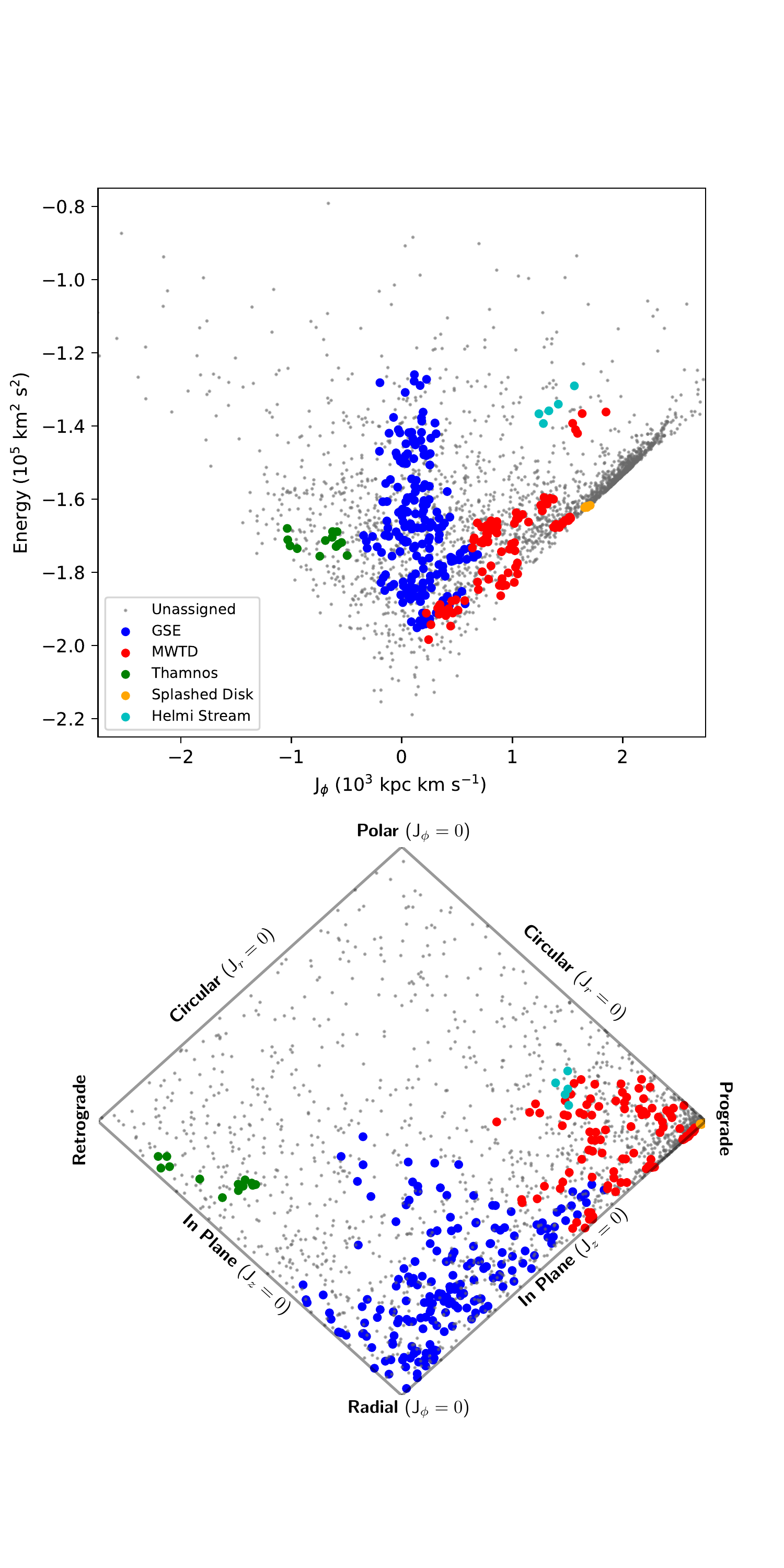}
    \caption{Top Panel: Lindblad Diagram of the identified MW substructures. The different structures are associated with the colors outlined in the legend. Bottom Panel: The projected-action plot of the same substructures. This space is represented by J$_{\phi}$/J$_{\text{Tot}}$ for the horizontal axis and (J$_{\text{z}}$ - J$_{\text{r}}$)/J$_{\text{Tot}}$ for the vertical axis with J$_{\text{Tot}}$ = J$_{\text{r}}$ + $|$J$_{\phi}|$ + J$_{\text{z}}$. For more details on the projected-action space, see Figure~3.25 in \citet{Binney2008}.}
    \label{fig:energy_actions}
\end{figure*}

Analyzing the orbital energy and actions is not sufficient to determine separate large-scale substructures for a variety of reasons. For example, the identified substructures in energy and action space can be overlapping; other dynamical parameters such as the eccentricity and the angular momentum can be used to resolve this degeneracy. Furthermore, the chemical properties of the stars can help to distinguish these substructures, through both the metallicity and the $\alpha$-element abundance. This information is crucial due to the differing star-formation histories of the structures, which can vary in both mass and formation redshift \citep{Naidu2020}. The outline for the prescription used to determine the structural associations with our DTGs is described in \citet{Naidu2020}. These authors use the Hectochelle in the Halo at High-resolution survey (H3; \citealt{Conroy2019}) to determine high-latitude substructure in the MW. Simple selections are performed based on physically motivated choices for each substructure, excluding previously defined substructures, as the process iterates to decrease contamination between substructures. See Section 3.2 in \citealt{Naidu2020} for a detailed description of the substructure determination procedures. Following their procedures, we find $5$ predominant MW substructures associated with our DTGs, listed in Table~\ref{tab:substructures}. This table provides the numbers of stars in each substructure, the mean and dispersion of their chemical abundances, and the mean and dispersion of their dynamical parameters for each substructure.  The Lindblad diagram and projected-action plot for these substructures is shown in Figure~\ref{fig:energy_actions}.

\input{Tables/substructure_table}

\subsubsection{Gaia-Sausage-Enceladus}\label{subsubsec:GSE}

The most populated substructure is Gaia-Sausage-Enceladus (GSE), which contains $186$ member stars. GSE is thought to be a remnant of an earlier merger that distributed a significant number of stars throughout the inner halo of the MW \citep{Belokurov2018,Helmi2018}. The action space determined by the member stars shows an extended radial component, a null azimuthal component within errors, and a null vertical component. These orbital properties are the product of the high-eccentricity selection of the DTGs, and agree with previous findings of GSE orbital characteristics when using other selection criteria \citep{Koppelman2018,Myeong2018b,Limberg2021a}. 

The $\langle$[Fe/H]$\rangle$ of GSE found in our work is very metal-poor ([Fe/H] $\leq - 2.0$), consistent with studies of its metallicity in dynamical groupings, even though our sample contains more metal-rich stars that could have been associated with GSE \citep{Gudin2021,Limberg2021a}. The stars that form DTGs in GSE tend to favor the more metal-poor tail of the substructure, which is also seen in previous DTG analysis. The $\langle$[$\alpha$/Fe]$\rangle$ of GSE exhibits a relatively low level, consistent with the low-Mg structure detected by \citet{Hayes2018} and with Mg levels consistent with accreted structures simulated by \citet{Mackereth2019}. The $\alpha$-element enhancement seen in GSE is due to accretion of older stellar populations, consistent with known element abundance patterns. We also obtain a $\langle$[C/Fe]$_{\textit{c}}\rangle$ for GSE (note that stars from \citealt{Huang2021c} do not have measurements of [C/Fe]$_{\textit{c}}$ or [$\alpha$/Fe]). Finally, we can associate the globular clusters UKS 1, Ryu 879 (RLGC2), and NGC 6284 with GSE, based on DTGs with similar orbital characteristics of these globular clusters (see Sec. \ref{subsec:GCDG} for details). Note in Figure~\ref{fig:energy_actions} how GSE occupies a large region of the Lindblad diagram, concentrated in the planar and radial portions of the projected-action plot.

\subsubsection{The Metal-Weak Thick Disk}\label{subsubsec:MWTD}

The second-most populated substructure is the Metal-Weak Thick Disk (MWTD), which contains $105$ member stars. The MWTD is thought to be due to either a merger scenario, possibly related to GSE, or the result of old stars born within the Solar radius migrating out to the Solar position due to tidal instabilities within the MW \citep{Carollo2019}. The non-existent radial and vertical velocity components, as well as the large positive azimuthal velocity component of the MWTD are all consistent with the velocity distribution for the MWTD from Carollo et al., even with the [Fe/H] cut containing more metal-rich stars than in their sample. The mean eccentricity distribution found within this substructure is also similar to that reported by Carollo et al., showing that the MWTD is a distinct component from the canonical thick disk (TD). Recently, both \citet{An2020} and \citet{Dietz2021} have presented evidence that the MWTD is an independent structure from the TD. The distribution in $\langle$[Fe/H]$\rangle$ and mean velocity space represents a stellar population consistent with the high-Mg population \citep{Hayes2018}, with the mean $\alpha$-element abundance being similar within errors. The $\langle$[C/Fe]$_{\textit{c}}\rangle$ abundance is also given for the MWTD, and shows an enhancement in carbon, possibly pointing to a relation with the strongly prograde CEMP structure found in \citet{Dietz2021}, which was attributed to the MWTD population.  Notice in Figure~\ref{fig:energy_actions} how the MWTD occupies a higher energy component of the disk (the gray dots mostly positioned with prograde orbits) in the Lindblad diagram. There are some MWTD stars that are close to the Helmi Stream stars; this is a by-product of the selection criteria for the MWTD stars, which is also present in \citet{Naidu2020} (See their Figure~12 and Figure~16).

\subsubsection{Thamnos}\label{subsubsec:Thamnos}

The third-most populated substructure is Thamnos, which contains $13$ member stars. Thamnos was proposed by \citet{Koppelman2019a} as a merger event that populated these stars in a retrograde orbit that is similar to TD stars. The low energy and strong retrograde rotation suggest that Thamnos merged with the MW long ago \citep{Koppelman2019a}. Here we find a similar low mean orbital energy and strong mean retrograde motion, though we do not recover as strong a retrograde motion as in \citet{Koppelman2019a}. The low mean metallicity, consistent with the value reported by \citet{Limberg2021a}, and elevated $\langle$[C/Fe]$_{c}\rangle$ of these stars also supports the merger being ancient. The $\langle$[$\alpha$/Fe]$\rangle$ is high, also suggesting an old population, consistent with \citet{Kordopatis2020}. Notice in Figure~\ref{fig:energy_actions} how Thamnos occupies a space that could be described as a retrograde version of disk stars.  

\subsubsection{The Splashed Disk}\label{subsubsec:SD}

The fourth-most populated substructure is the Splashed Disk (SD), which contains $7$ member stars. The SD is thought to be a component of the primordial MW disk that was kinematically heated during the GSE merger event \citep{Helmi2018,DiMatteo2019,Belokurov2020}. The mean velocity components of the SD are consistent with a null radial and vertical velocity, while showing a large positive azimuthal velocity consistent with disk-like stars. The mean eccentricity of these stars is consistent with disk-like orbits. The SD consists of the most metal-rich substructure identified here. The high $\langle$[$\alpha$/Fe]$\rangle$ abundances for the SD shows that these stars are old, and they could be the result of a possible merger event, such as the one that created GSE. The $\langle$[C/Fe]$_{\textit{c}}\rangle$ abundance for the SD is sub-Solar, which is an interesting contrast to the high mean $\alpha$-element abundances. Notice in Figure~\ref{fig:energy_actions} how the SD overlaps with the MWTD. This is due to the selection criteria only using metallicity and $\alpha$-element abundances to determine the SD stars \citep{Naidu2020}. Considering the SD is thought to be composed of stars that have been heated due to the GSE merger event, the positions of the SD stars in the Lindblad diagram does not show a relatively large deviation from disk-like orbits. More associations with the SD are needed to make any definitive claims.

\subsubsection{The Helmi Stream}\label{subsubsec:Helmi}

The least populated substructure is the Helmi Stream (HS), which contains only $5$ member stars. The HS is one of the first detected dynamical substructures in the MW using integral of motions \citep{Helmi1999}. The HS has a characteristically high vertical velocity, which separates it from other stars that lie in the disk, and can be seen in the sample here. The large uncertainty on vertical velocity of the HS members corresponds to the positive and negative vertical velocity components of the stream, with the negative vertical velocity population dominating, consistent with the members determined here \citep{Helmi2020}. The $\langle$[Fe/H]$\rangle$ of the HS is more metal-poor in this sample, compared to the known HS members ([Fe/H] $\sim -1.5$; \citealt{Koppelman2019b}). Recently however, \citet{Limberg2021b} noted that the metallicity range of HS is more metal-poor than previously expected, with stars reaching down to [Fe/H] $\sim -2.5$, which is consistent with the results presented here. Notice in Figure~\ref{fig:energy_actions} how the HS occupies a relatively isolated space in the Lindblad diagram, thanks to the large vertical velocity of the stars providing the extra energy compared to the other disk stars.



\subsection{Previously Identified Dynamically Tagged Groups and Stellar Associations}\label{subsec:Prev_DTGs_Stellar_assoc}

\input{Tables/interesting_substructure_table}

Separately, we can compare the newly identified DTGs in this work with other dynamical groups identified by previous authors.  This provides sanity checks throughout the process by making sure associations are consistent across various works, even if the orbital calculations differ due to the use of slightly different Galactic models. We take the mean group properties used to detect the previously identified groups and compare them to the mean and dispersion for the dynamical parameters of our identified DTGs. Matches are recovered when all the dynamical parameters are within one sigma of each other. This is applied to every DTG discovered in the following papers, using the parameter space that was explored in their discovery \citep{Helmi2017,Myeong2017,Koppelman2018,Myeong2018b,Myeong2018c,Roederer2018,Li2019,Sestito2019,Yuan2019,Borsato2020,Li2020,Monty2020,Yuan2020a,Yuan2020b,Cordoni2021,Gudin2021,Kielty2021,Limberg2021a}. Stellar associations are also considered, allowing the identification of stars in our sample that belong to previously identified groups. If our star is within a $5 \arcsec$ radius of another star in a previously identified group, then they are considered an association. The resulting dynamical associations between our identified DTGs and previously identified groups (along with substructure and globular cluster associations, see Section \ref{subsec:GCDG}) are listed in Table \ref{tab:interesting_substructure}. Table~\ref{tab:cluster_results_stub} lists the individual stellar associations for each of our DTGs.

One example of associations of identified DTGs with past groups is  DTG-46. This DTG was associated with GSE through the procedure outlined in \citet{Naidu2020} (see Section~\ref{subsec:MWSubstructure} for more details). There was one star associated with this DTG through a $5 \arcsec$ radius search of the DTG member stars. This star belongs in ZY20b:DTG-14, and was associated with the GSE as well \citep{Yuan2020b}. DTG-46 is also dynamically associated with GL21:DTG-38 which was discovered by \citet{Limberg2021a} and determined by their procedure to be a part of GSE, again consistent with the findings of DTG-46. The globular cluster UKS 1 is also dynamically associated with DTG-46, and is an old system which belongs to the bulge of the MW \citep{Fernandez-Trincado2020}, possibly indicating an origin that corresponds to the GSE merger event. 

Interesting DTGs associated with the GSE structure are DTG-7,8,10,11,18,22,27,28,33,42, and 46,  which have multiple previously identified groups and stars associated with them. Taking a closer look at DTG-7 and DTG-8, they both have the same associations -- GM17:Comoving, SM20:Sausage, and GC21:Sausage, all of which are identified as a part of GSE \citep{Myeong2017,Monty2020,Cordoni2021}. With these associations being the same for both DTG-7 and DTG-8, this may show similarities in the possible origin of these two DTGs. The mean chemical abundances of both these DTGs agree as well ($\langle$[Fe/H]$\rangle$ $\sim -2.1$, $\langle$[C/Fe]$_{c}\rangle \sim +0.4$, and $\langle$[$\alpha$/Fe]$\rangle$ $\sim +0.3$), further suggesting a common origin.

The only DTG with multiple associations related to the MWTD is DTG-2, which is associated with HL19:GL-1, DG21:CDTG-8, and DG21:CDTG-6.  Only DG21:CDTG-8 was associated with the MWTD by the authors, and the other two groups were unidentified \citep{Li2019,Gudin2021}. DTG-51 is an interesting case, since we have associated it with Thamnos, and there are four previously identified DTGs associated with the cluster (along with a stellar association from one of the associated DTGs). GL21:DTG-31 was associated with Thamnos \citep{Limberg2021a}, while HL19:GL-4 was unassigned \citep{Li2019}. However, both SM20:SeqG1 and GC21:Sequoia \citep{Monty2020,Cordoni2021} were previously identified as belonging to Sequoia \citep{Myeong2019}, which has a higher energy compared to Thamnos, though both are retrograde structures. \citet{Monty2020} made the decision to keep SM20:SeqG1 assigned to Sequoia rather than Thamnos, based on the possibility of Thamnos not being known as a separate structure from Sequoia at the time of their work, so they considered the association.  

Finally, there are unassigned MW substructure DTGs that have multiple associations -- DTG-4 and DTG-34.  DTG-4 has one stellar association and two previously identified groups associated with the cluster, though the stellar association's previously identified DTG did not fall in the group association. The stellar association was with GC21:Sausage \citep{Cordoni2021}, while the previously identified group associations were with DG21:CDTG-14 and GL21:DTG-13 \citep{Gudin2021,Limberg2021a}. Obviously, GC21:Sausage was identified as belonging to GSE by the authors, but both \citet{Gudin2021} and \citet{Limberg2021a} had their group unassigned to any larger substructure. GL21:DTG-13 was associated with ZY20b:DTG-39 (unassigned by \citealt{Yuan2020b}), though we do not recover that association \citep{Limberg2021a}. As another unassigned MW substructure cluster, DTG-34 has two previously identified group associations, GC21:Sequoia and GL21:DTG-22 \citep{Cordoni2021,Limberg2021a}, along with a globular cluster association of Ryu 879 (RLGC 2) \citep{Vasiliev2021}.  GC21:Sequoia was previously identified as belonging to Sequoia \citep{Cordoni2021}, while GL21:DTG-22 is again unassigned, but they did find an association to ZY20b:DTG-33 (also unassigned by \citealt{Yuan2020b}), which we did not recover. The DTGs that have unassigned MW substructure can offer valuable insights into the smaller structures of the MW that have yet to be confirmed.

Another use of the stellar associations comes from the suggestion by \citet{Roederer2018}, strengthened by \citet{Gudin2021}, that dynamical groups of stars have a statistically significant correlation between their elemental abundances. This is of importance to discover new chemically peculiar stars, particularly  $r$-process-enhanced stars. Out of our DTGs, there are $6$ associations between known $r$-process-enhanced stars and DTG-2,4,5,18,22, and 39. These DTGs provide interesting candidates for high-resolution spectroscopic follow-up, due to the increased likelihood of the other members comprising chemically peculiar stars, especially in terms of $r$-process enhancement \citep{Roederer2018,Gudin2021}. 

\subsection{Globular Clusters and Dwarf Galaxies}\label{subsec:GCDG}

Both globular clusters and dwarf galaxies have been shown to play an important role in the formation of chemically peculiar stars \citep{Ji2016,Myeong2018a}. Globular clusters can also be a good indicator of 
galaxy-formation history based on their metallicities and orbits \citep{Woody2021}. From the work of \citet{Vasiliev2021}, we can compare the dynamical properties of $170$ globular clusters to those of the DTGs we identify. The procedure that is employed is the same one used for previously identified groups and stellar associations introduced in Sec. \ref{subsec:Prev_DTGs_Stellar_assoc}. The globular clusters analyzed by \citet{Vasiliev2021} have orbits that are determined using the same potential, \MWMMXVII, and integrator, \AGAMA, as our procedure. The dynamics for $45$ dwarf galaxies of the MW (excluding the Large Magellanic Cloud, Small Magellanic Cloud, and Sagittarius) undergo the same orbit determination as our stars (See Sec. \ref{sec:DynamicalParameters}), based on the 6$-$D astrometric parameters from both \citet{McConnachie2020} and \citet{Li2021}. The same procedure used for previously identified groups was then applied to determine whether a DTG was dynamically associated to the dwarf galaxy. Stellar associations were also determined for both globular clusters and dwarf galaxies in the same manner as previously identified groups. 

The above comparison exercise led to $8$ globular cluster associations, with $5$ being unique. For a breakdown of which globular clusters are associated with our DTGs, see Table \ref{tab:interesting_substructure}. Ryu 879 (RLGC 2) has three DTG associations which agree with each other in mean metallicity ($\langle$[Fe/H]$\rangle$ $\sim -2.4$), mean carbon abundance ($\langle$[C/Fe]$_{c}\rangle \sim +0.35$), and mean $\alpha$-element abundance ($\langle$[$\alpha$/Fe]$\rangle$ $\sim +0.3$). The mean metallicity agrees with the discovery of the globular cluster Ryu 879 (RLGC 2) \citep{Ryu2018}; carbon and $\alpha$-element abundance estimates are presented here for the first time. NGC 5139 ($\omega$ Cen) has two DTG associations which agree with each other in mean metallicity ($\langle$[Fe/H]$\rangle$ $\sim -2.0$) and $\alpha$-element abundances ($\langle$[$\alpha$/Fe]$\rangle$ $\sim +0.2$), but not in mean carbon abundances ($\langle$[C/Fe]$_{c}\rangle \sim +0.7$ vs. $\langle$[C/Fe]$_{c}\rangle \sim +0.3$). The other globular clusters with matched dynamics in this work are NGC 6284 and UKS-1, while NGC 6397 had a stellar association. Even though the star in NGC 6397 would have individually been associated with the globular cluster orbital parameters, the overall DTG did not possess sufficiently similar orbital characteristics to be associated. The DTGs associated with globular clusters are expected to have formed in chemically similar birth environments, and is mostly supported through the similar chemical properties of the DTGs. Associations of globular clusters with Galactic substructure have been made by \cite{Massari2019}. These authors did not analyze Ryu 879 (RLGC 2), since the globular cluster was recently discovered at the time of the publication. NGC 5139 ($\omega$ Cen) was identified as being associated with GSE or Sequoia by the authors, but our associations did not recover this match. NGC 6284 was found to be associated to GSE by both our identification and \cite{Massari2019}. Our association found that UKS-1 is part of GSE, while the authors could not identify the origin. NGC 6397 is not associated to any substructure by our procedure, but \cite{Massari2019} find it associated to the main disk, which we do not consider as part of the substructure routine.

We did not identify any associations of DTGs to the sample of (surviving) MW dwarf galaxies, either through stellar associations, or through the dynamical association procedure described above. Nevertheless, some of the DTGs identified by our analysis may well be associated with dwarf galaxies that have previously merged with the MW.

\section{Global Properties of Identified DTGs}\label{sec:Global}

\begin{figure}[t]
    \includegraphics[width=0.48\textwidth,height=0.48\textheight]{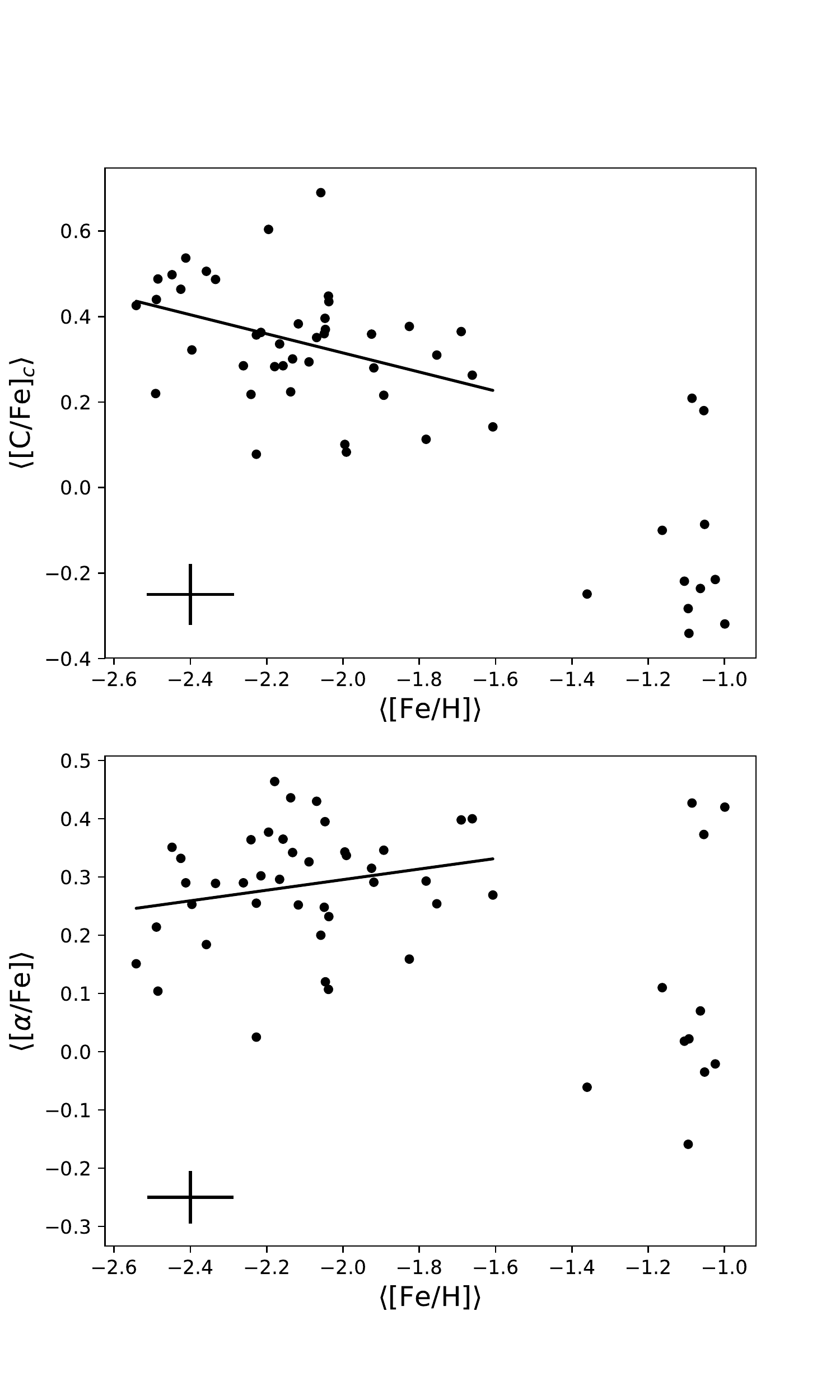}
    \caption{Top Panel: The mean corrected carbon abundances ($\langle$[C/Fe]$_{c}\rangle$) for the identified DTGs. Bottom Panel: The mean $\alpha$-element abundance ($\langle$[$\alpha$/Fe]$\rangle$) for the identified DTGs. Both Panels: The typical error of the mean for the DTGs is indicated with the black cross. Trend lines for the metal-poor DTGs ([Fe/H] $\leq -1.6$) are shown as the black line.}
    \label{fig:cluster_abundances}
\end{figure}

\begin{figure*}[t]
    \centering
    \includegraphics[width=0.98\textwidth,keepaspectratio]{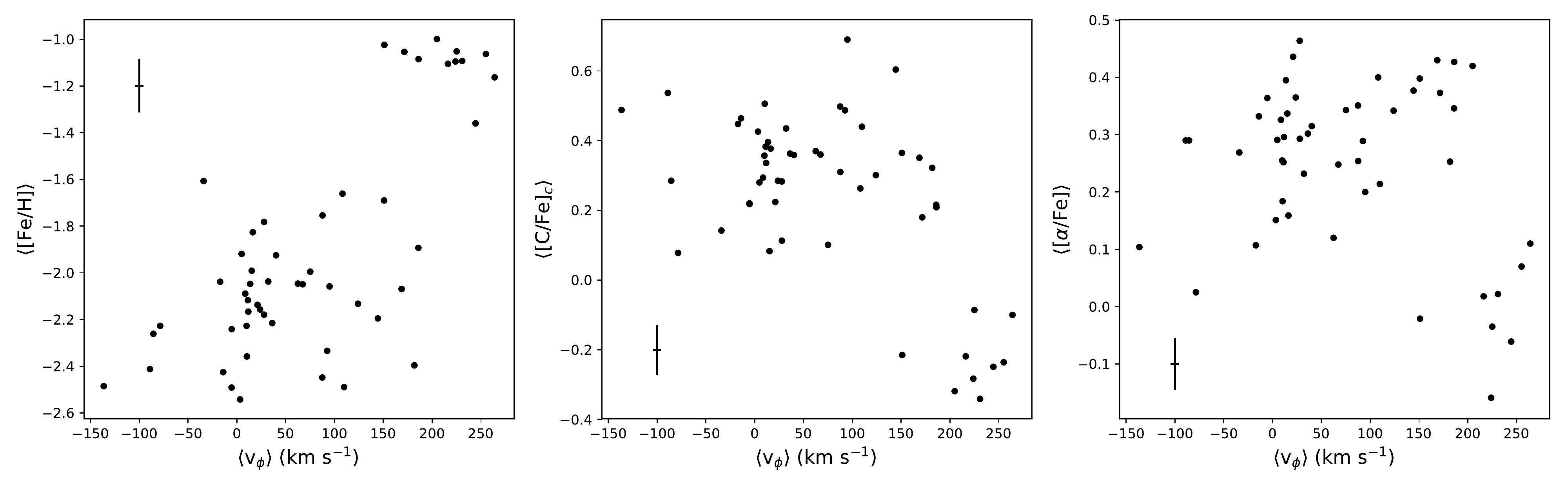}
    \caption{The mean metallicity ($\langle$[Fe/H]$\rangle$, left), mean carbon ($\langle$[C/Fe]$_{c}\rangle$, middle) and mean $\alpha$-element ($\langle$[$\alpha$/Fe]$\rangle$, right) abundances of the DTGs, as functions of the mean azimuthal velocity ($\langle$v$_{\phi}\rangle$). The typical error of the mean for the DTGs is identified with the black crosses.}
    \label{fig:clust_orbit_abundances}
\end{figure*}

We now consider the chemical and dynamical behaviors of the 52 identified DTGs in this work as a whole. Figure~\ref{fig:cluster_abundances} shows the biweight location and a representative error of the mean for each of the DTGs in the chemical spaces of $\langle$[C/Fe]$_{c}\rangle$ and $\langle$[$\alpha$/Fe]$\rangle$, as a function of $\langle$[Fe/H]$\rangle$, where available. Interestingly, there appears to be a distinction between DTGs that are more metal-poor ($\langle$[Fe/H]$\rangle$ $\leq -1.6$) compared to those that are more metal-rich ($\langle$[Fe/H]$\rangle$ $\sim -1.1$). The mean carbon abundance is consistently higher in DTGs that are more metal poor. The mean $\alpha$-element abundances are higher for DTGs that are part of the metal-poor regime, and tend to increase in mean $\alpha$-element abundance as the mean metallicity increases; there is no clear trend in the metal-rich regime. If these DTGs were populated by random stars that had abundances chosen from the Initial sample, the trends that we observe across both the mean carbon and $\alpha$-element abundances would not appear.

Next, we consider how the mean metallicities and mean elemental abundances of the DTGs are distributed over the mean azimuthal velocity, as shown in Figure~\ref{fig:clust_orbit_abundances}. The azimuthal velocity does not depend on the adopted potential, and thus provides an independent check on the chemistry from the potential-dependant clusters. In the panels of Figure~\ref{fig:clust_orbit_abundances}, one can see that the DTGs are split into the prograde structures that are part of the disk populations and the other structures that span the rest of the MW, including a majority of the halo DTGs identified with GSE. The chemistry appears distinct as well. For mean metallicity ($\langle$[Fe/H]$\rangle$, left panel), it is clear that the disk-like stars are associated with more metal-rich compositions while the rest of the sample is spread out across the more metal-poor portion of the sample. The mean carbon abundances ($\langle$[C/Fe]$_{c}\rangle$, middle panel) exhibit lower carbon abundances as compared to their halo-like counterparts, commensurate with the known distributions of mean carbon vs. mean metallicity shown in Figure~\ref{fig:cluster_abundances}. The mean $\alpha$-element abundances ($\langle$[$\alpha$/Fe]$\rangle$, right panel) show another distinction between the disk-like DTGs and the halo-dominated component. There exists a possible trend in the mean $\alpha$-element abundances appearing in the halo-dominated component as well, possibly suggesting that the retrograde structures are younger than the prograde structures, since the mean $\alpha$-element abundance is expected to decline with prolonged episodes of star formation.  It will be of interest to further examine these trends with larger samples of DTGs identified from additional analyses of field stars in the near future.

\section{Discussion}\label{sec:Discussion}

We have assembled an Initial Sample of 6445 stars from the Best $\&$ Brightest survey \citep{Schlaufman2014} with available estimates of [Fe/H], and in some cases, with [C/Fe] and [$\alpha$/Fe]. The Final Sample of $2526$ metal-poor stars ([Fe/H] $\leq -0.8$) had sufficient information with which orbits were constructed, in order to determine Dynamically Tagged Groups (DTGs) in orbital energy and cylindrical action space with the \HDBSCAN ~algorithm. We chose \HDBSCAN\ as the clustering algorithm due to precedence within the literature, \citep{Koppelman2019a,Gudin2021,Limberg2021a} and its ability to extract clusters of stars over the energy and action space. Other clustering algorithms have been considered in the past, such as agglomerative clustering, affinity propogation, K-means, and mean-shift clustering \citep{Roederer2018}, along with friends-of-friends \citep{Gudin2021}. 

We recover $52$ DTGs that include between $5$ and $21$ members, with $18$ DTGs containing at least $10$ member stars. These DTGs were associated with MW substructure, resulting in the identification of the Gaia-Sausage-Enceladus, the Metal-Weak Thick Disk, Thamnos, the Splashed Disk, and the Helmi Stream. A total of $5$ unique Globular Clusters were associated with $8$ different DTGs, while no surviving Dwarf Galaxies were determined to be associated with the identified DTGs. Previously identified groups were found to be associated with the DTGs as well, with past work mostly confirming our substructure identification. Each of these associations allow insights into the dynamical and chemical properties of the parent substructures. 

The implications of past group and stellar associations were explored with emphasis placed on the structure associations. Chemically peculiar stellar associations and previously identified Chemo-Dynamically Tagged Groups (CDTGs) were addressed as being good candidates for high-resolution follow-up spectroscopy targets, due to the statistical likelihood of the other members being chemically peculiar as well, mostly focused on $r$-process-enhanced stars. 

Comparing the DTGs mean metallicities, carbon, and $\alpha$-element abundances showed that these DTGs can be separated into metal-rich disk-like and metal-poor halo-like components. Analyzing the orbital characteristics of the mean azimuthal velocity, it was determined that the DTGs follow expected trends in the metallicities and abundances over the range of these parameters. 

The methods presented here will be used on larger samples of field stars that we are in the process of assembling -- both RAVE DR6 \citep{Steinmetz2020} and the HK/HES/HKII surveys \citep{Beers1985,Beers1992,Christlieb2008,Rhee2001} (a subset of which were analyzed by \citealt{Limberg2021a}). These data sets will also be supplemented with photometric estimages of effective temperature and metallicity from \citet{Huang2021c}, and, for instance, allows stars from the HK/HES/HKII surveys with no previous spectroscopic follow-up to be explored, expanding the data set used by \citet{Limberg2021a}. 

\vspace{2.0cm}

The authors are thankful to an anonymous referee, who provided comments and suggestions that substantially improved this paper. 
We thank K.C.S and A.R.C for kindly providing early access to the candidate lists from the Best $\&$ Brightest Survey. D.S. and T.C.B. acknowledge partial support for this work from grant PHY 14-30152; Physics Frontier Center/JINA Center for the Evolution of the Elements (JINA-CEE), awarded by the US National Science Foundation. The work of V.M.P. is supported by NOIRLab, which is managed by the Association of Universities for Research in Astronomy (AURA) under a cooperative agreement with the National Science Foundation. G.L. acknowledges CAPES (PROEX; Proc. 88887.481172/2020-00). Z.Y. acknowledges funding from the Agence Nationale de la Recherche (ANR project ANR-18-CE31-0017) and the European Research Council (ERC) under the European Unions Horizon 2020 research and innovation programme (grant agreement No. 834148). Y.S.L. acknowledges support from the National Research Foundation (NRF) of Korea grant funded by the Ministry of Science and ICT (NRF-2021R1A2C1008679). R.M.S. acknowledges CNPq (Proc. 436696/2018-5 and 306667/2020-7). A.R.C. is supported in part by the Australian Research Council through a Discovery Early Career Researcher Award (DE190100656). Parts of this research were supported by the Australian Research Council Centre of Excellence for All Sky Astrophysics in 3 Dimensions (ASTRO 3D), through project number CE170100013. K.H. is supported by JSPS KAKENHI Grant Numbers JP21K13965 and JP21H00053.

\section{Appendix}

Here we present the tables for the Initial and Final Samples of the Best $\&$ Brightest survey.

\include{Tables/full_data_description_table}

\include{Tables/final_data_description_table}

\bibliography{main}{}
\bibliographystyle{aasjournal}

\end{document}

%% file: Tables/cluster_summary_table.tex
\begin{deluxetable*}{l  r  r  c}
\tabletypesize{\scriptsize}
\tablecaption{Identified DTGs \label{tab:cluster_summary}}
\tablehead{\colhead{DTG} & \colhead{$N$ Stars} & \colhead{Confidence} & \colhead{Associations}}
\startdata
1 & $21$ & $77.1\%$ & new\\
2 & $20$ & $72.9\%$ & MWTD, DG21:CDTG-8, HL19:GL-1, DG21:CDTG-6\\
3 & $18$ & $91.7\%$ & GSE, GC21:Sausage\\
4 & $18$ & $91.5\%$ & GC21:Sausage, DG21:CDTG-14, GL21:DTG-13\\
5 & $17$ & $100.0\%$ & GSE, DG21:CDTG-22\\
6 & $15$ & $92.9\%$ & ZY20b:DTG-35\\
7 & $13$ & $84.2\%$ & GSE, GM17:Comoving, SM20:Sausage, GC21:Sausage\\
8 & $12$ & $57.8\%$ & GSE, GM17:Comoving, SM20:Sausage, GC21:Sausage\\
9 & $12$ & $100.0\%$ & EV21:NGC~6397\\
10 & $12$ & $96.1\%$ & GSE, GC21:Sausage, GL21:DTG-23\\
11 & $12$ & $63.6\%$ & GSE, GM17:Comoving, GC21:Sausage, EV21:Ryu~879~(RLGC~2)\\
12 & $11$ & $59.8\%$ & MWTD, FS19:P\\
13 & $11$ & $63.1\%$ & new\\
14 & $11$ & $100.0\%$ & MWTD\\
15 & $10$ & $95.4\%$ & MWTD\\
16 & $10$ & $99.9\%$ & GSE\\
17 & $9$ & $92.5\%$ & GSE, GC21:Sausage\\
18 & $9$ & $85.5\%$ & GSE, DG21:CDTG-13, GC21:Sausage, GL21:DTG-30\\
19 & $9$ & $62.9\%$ & MWTD\\
20 & $8$ & $100.0\%$ & MWTD\\
21 & $8$ & $94.1\%$ & GSE\\
22 & $8$ & $85.1\%$ & GSE, DG21:CDTG-1, GL21:DTG-30\\
23 & $8$ & $59.4\%$ & new\\
24 & $8$ & $65.7\%$ & MWTD\\
25 & $8$ & $81.7\%$ & MWTD, HL19:GL-1\\
26 & $8$ & $89.7\%$ & Thamnos\\
27 & $8$ & $53.9\%$ & GSE, SM20:Sausage, GC21:Sausage\\
28 & $7$ & $84.0\%$ & GSE, GM17:Comoving, GC21:Sausage, EV21:Ryu~879~(RLGC~2)\\
29 & $7$ & $93.4\%$ & GSE\\
30 & $7$ & $62.7\%$ & new\\
31 & $7$ & $95.5\%$ & Splashed Disk\\
32 & $7$ & $49.3\%$ & new\\
33 & $6$ & $67.5\%$ & GSE, GC21:Sausage, EV21:NGC~6284\\
34 & $6$ & $53.2\%$ & GC21:Sequoia, GL21:DTG-22, EV21:Ryu~879~(RLGC~2)\\
35 & $6$ & $72.2\%$ & EV21:NGC~5139~(oCen)\\
36 & $6$ & $63.3\%$ & GSE\\
37 & $6$ & $70.2\%$ & new\\
38 & $6$ & $61.9\%$ & new\\
39 & $6$ & $81.5\%$ & DG21:CDTG-14\\
40 & $6$ & $97.9\%$ & MWTD\\
41 & $6$ & $59.7\%$ & GSE\\
42 & $6$ & $100.0\%$ & GSE, SM20:Sausage, GC21:Sausage\\
43 & $5$ & $99.8\%$ & GSE\\
44 & $5$ & $78.6\%$ & MWTD\\
45 & $5$ & $61.5\%$ & MWTD, GL21:DTG-10\\
46 & $5$ & $99.9\%$ & GSE, ZY20b:DTG-14, GL21:DTG-38, EV21:UKS~1\\
47 & $5$ & $36.5\%$ & new\\
48 & $5$ & $42.4\%$ & new\\
49 & $5$ & $99.8\%$ & MWTD\\
50 & $5$ & $72.8\%$ & EV21:NGC~5139~(oCen)\\
51 & $5$ & $99.8\%$ & Thamnos, GC21:Sequoia, HL19:GL-4, SM20:SeqG1, GL21:DTG-31\\
52 & $5$ & $78.8\%$ & Helmi Stream\\
\enddata
\tablecomments{We adopt the nomenclature for previously identified DTGs and CDTGs from \cite{Yuan2020b}.}\end{deluxetable*}

%% file: Tables/cluster_stellar_results_stub_table.tex
\begin{deluxetable*}{l  c  c  c  c}
\tablecaption{DTGs Identified by \HDBSCAN \label{tab:cluster_results_stub}}
\tablehead{\colhead{Star Name} & \colhead{[Fe/H]} & \colhead{[C/Fe]} & \colhead{[C/Fe]$_{c}$} & \colhead{[$\alpha$/Fe]}}
\startdata
\multicolumn{5}{c}{$DTG-1$} \\
\multicolumn{5}{c}{Structure: Unassigned Structure} \\
\multicolumn{5}{c}{Group Assoc: No Group Associations} \\
\multicolumn{5}{c}{Stellar Assoc: No Stellar Associations} \\
\multicolumn{5}{c}{Globular Assoc: No Globular Associations} \\
\multicolumn{5}{c}{Dwarf Galaxy Assoc: No Dwarf Galaxy Associations} \\
J035806.92$-$412551.4 & $-1.061$ & $\dots$ & $\dots$ & $\dots$ \\
J060951.09$-$233050.3 & $-0.816$ & $\dots$ & $\dots$ & $\dots$ \\
J062203.79$-$215529.0 & $-1.034$ & $+0.094$ & $+0.094$ & $\dots$ \\
J075523.99$+$352005.5 & $-0.927$ & $-0.273$ & $-0.273$ & $+0.018$ \\
J095327.59$-$271710.2 & $-1.430$ & $\dots$ & $\dots$ & $\dots$ \\
J102813.29$-$294653.2 & $-1.231$ & $\dots$ & $\dots$ & $\dots$ \\
J114658.41$+$483515.4 & $-1.172$ & $-0.340$ & $-0.340$ & $-0.029$ \\
J115951.48$-$390527.5 & $-1.041$ & $+0.077$ & $+0.077$ & $+0.167$ \\
J133023.82$-$000654.8 & $-0.878$ & $\dots$ & $\dots$ & $\dots$ \\
J155713.55$-$375241.2 & $-1.169$ & $-0.216$ & $-0.186$ & $\dots$ \\
J160749.07$+$142013.2 & $-1.468$ & $-0.208$ & $-0.208$ & $-0.007$ \\
J161730.55$-$193225.6 & $-1.094$ & $+0.402$ & $+0.422$ & $-0.035$ \\
J173313.27$-$610148.2 & $-1.359$ & $-0.207$ & $-0.207$ & $\dots$ \\
J173346.78$-$674403.0 & $-1.658$ & $\dots$ & $\dots$ & $\dots$ \\
J183718.56$-$641756.9 & $-0.871$ & $\dots$ & $\dots$ & $\dots$ \\
J195013.73$-$771450.7 & $-1.102$ & $-0.558$ & $-0.558$ & $-0.275$ \\
J202034.71$-$275711.3 & $-1.083$ & $-0.143$ & $-0.143$ & $+0.049$ \\
J205902.49$+$184702.0 & $-1.382$ & $-0.268$ & $-0.258$ & $\dots$ \\
J220339.84$-$644528.9 & $-0.872$ & $\dots$ & $\dots$ & $\dots$ \\
J222900.16$-$173756.1 & $-1.108$ & $\dots$ & $\dots$ & $\dots$ \\
J223557.26$-$243411.6 & $-0.945$ & $-0.265$ & $-0.265$ & $+0.062$ \\
$\mu \pm \sigma$ ([X/Y]) & $-1.105\pm0.222$ & $-0.225\pm0.169$ & $-0.219\pm0.170$ & $+0.018\pm0.086$\\
\pagebreak
\enddata
\tablecomments{$\mu$ and $\sigma$ represent the biweight estimates of the location and scale for the abundances in the DTG.}\tablecomments{This table is a stub; the full table is available in the electronic edition.}\end{deluxetable*}

%% file: Tables/cluster_orbital_stub_table.tex
\begin{deluxetable*}{l  r  c  c  c  r}
\tablecaption{Cluster Dynamical Parameters Determined by \AGAMA \label{tab:cluster_orbit_stub}}
\tablehead{\colhead{Cluster} & \colhead{$N$ Stars} & \colhead{($\langle$v$_{\text{r}}\rangle$,$\langle$v$_{\phi}\rangle$,$\langle$v$_{\text{z}}\rangle$)} & \colhead{($\langle$J$_{\text{r}}\rangle$,$\langle$J$_{\phi}\rangle$,$\langle$J$_{\text{z}}\rangle$)} & \colhead{$\langle$E$\rangle$} & \colhead{$\langle$ecc$\rangle$}\\
\colhead{} & \colhead{} & \colhead{($\sigma_{\langle\text{v}_{\text{r}}\rangle}$,$\sigma_{\langle\text{v}_{\phi}\rangle}$,$\sigma_{\langle\text{v}_{\text{z}}\rangle}$)} & \colhead{($\sigma_{\langle\text{J}_{\text{r}}\rangle}$,$\sigma_{\langle\text{J}_{\phi}\rangle}$,$\sigma_{\langle\text{J}_{\text{z}}\rangle}$)} & \colhead{$\sigma_{\langle\text{E}\rangle}$} & \colhead{$\sigma_{\langle\text{ecc}\rangle}$}\\
\colhead{} & \colhead{} & \colhead{(km s$^{-1}$)} & \colhead{(kpc km s$^{-1}$)} & \colhead{(10$^{5}$ km$^{2}$ s$^{-2}$)} & \colhead{}}
\startdata
$DTG-1$ & $21$ & ($0.8$,$216.2$,$-1.0$) & ($10.3$,$1749.2$,$2.3$) & $-1.603$ & $0.090$ \\
 & & ($21.3$,$7.4$,$15.0$) & ($5.9$,$29.8$,$2.4$) &  \textcolor{white}{+}$0.007$ & $0.026$ \\
$DTG-2$ & $20$ & ($12.7$,$124.0$,$-21.8$) & ($163.5$,$936.0$,$273.8$) & $-1.664$ & $0.421$ \\
 & & ($90.4$,$37.1$,$106.0$) & ($75.7$,$219.9$,$26.7$) &  \textcolor{white}{+}$0.036$ & $0.113$ \\
$DTG-3$ & $18$ & ($23.7$,$4.8$,$0.5$) & ($894.7$,$45.8$,$58.5$) & $-1.644$ & $0.932$ \\
 & & ($153.2$,$13.0$,$38.5$) & ($48.5$,$120.4$,$17.9$) &  \textcolor{white}{+}$0.028$ & $0.040$ \\
$DTG-4$ & $18$ & ($-22.4$,$-14.0$,$-81.0$) & ($159.5$,$-80.7$,$1032.0$) & $-1.661$ & $0.443$ \\
 & & ($133.1$,$39.9$,$143.3$) & ($69.7$,$182.7$,$96.0$) &  \textcolor{white}{+}$0.052$ & $0.104$ \\
$DTG-5$ & $17$ & ($-40.7$,$67.4$,$-4.6$) & ($447.1$,$492.3$,$69.5$) & $-1.765$ & $0.733$ \\
 & & ($103.8$,$21.2$,$58.2$) & ($59.9$,$118.8$,$17.6$) &  \textcolor{white}{+}$0.022$ & $0.062$ \\
\enddata
\tablecomments{This table is a stub; the full table is available in the electronic edition.}\end{deluxetable*}

%% file: Tables/substructure_table.tex
\begin{deluxetable*}{l  r  r  r  r  c  c  c  r}
\tablecaption{Identified Milky Way Substructures \label{tab:substructures}}
\tablehead{\colhead{MW Substructure} & \colhead{$N$ Stars} & \colhead{$\langle$[Fe/H]$\rangle$} & \colhead{$\langle$[C/Fe]$_{\textit{c}}\rangle$} & \colhead{$\langle$[$\alpha$/Fe]$\rangle$} & \colhead{($\langle$v$_{\text{r}}\rangle$,$\langle$v$_{\phi}\rangle$,$\langle$v$_{\text{z}}\rangle$)} & \colhead{($\langle$J$_{\text{r}}\rangle$,$\langle$J$_{\phi}\rangle$,$\langle$J$_{\text{z}}\rangle$)} & \colhead{$\langle$E$\rangle$} & \colhead{$\langle$ecc$\rangle$}\\
\colhead{} & \colhead{} & \colhead{} & \colhead{} & \colhead{} & \colhead{($\sigma_{\langle\text{v}_{\text{r}}\rangle}$,$\sigma_{\langle\text{v}_{\phi}\rangle}$,$\sigma_{\langle\text{v}_{\text{z}}\rangle}$)} & \colhead{($\sigma_{\langle\text{J}_{\text{r}}\rangle}$,$\sigma_{\langle\text{J}_{\phi}\rangle}$,$\sigma_{\langle\text{J}_{\text{z}}\rangle}$)} & \colhead{$\sigma_{\langle\text{E}\rangle}$} & \colhead{$\sigma_{\langle\text{ecc}\rangle}$}\\
\colhead{} & \colhead{} & \colhead{} & \colhead{} & \colhead{} & \colhead{(km s$^{-1}$)} & \colhead{(kpc km s$^{-1}$)} & \colhead{(10$^{5}$ km$^{2}$ s$^{-2}$)} & \colhead{}}
\startdata
GSE & 184 & $-2.067$ & $+0.316$ & $+0.290$ & ($-17.9$,$20.8$,$3.6$) & ($785.1$,$146.0$,$122.1$) & $-1.675$ & $0.887$\\
 &  & $0.413$ & $0.257$ & $0.185$ & ($164.0$,$27.8$,$69.7$) & ($386.8$,$199.1$,$95.1$) & \textcolor{white}{+}$0.166$ & $0.087$\\
\hline
MWTD & 106 & $-1.914$ & $+0.321$ & $+0.339$ & ($4.8$,$136.7$,$1.6$) & ($204.4$,$964.6$,$116.6$) & $-1.722$ & $0.455$\\
 &  & $0.601$ & $0.278$ & $0.168$ & ($91.5$,$45.8$,$64.2$) & ($146.8$,$385.5$,$101.1$) & \textcolor{white}{+}$0.122$ & $0.166$\\
\hline
Thamnos & 13 & $-2.256$ & $+0.457$ & $+0.269$ & ($43.2$,$-104.1$,$-10.4$) & ($346.1$,$-731.7$,$102.5$) & $-1.718$ & $0.623$\\
 &  & $0.527$ & $0.249$ & $0.141$ & ($120.6$,$29.0$,$51.8$) & ($80.7$,$193.5$,$48.3$) & \textcolor{white}{+}$0.023$ & $0.094$\\
\hline
Splashed Disk & 7 & $-1.002$ & $-0.319$ & $+0.420$ & ($-1.6$,$207.5$,$6.8$) & ($20.4$,$1674.7$,$3.1$) & $-1.621$ & $0.129$\\
 &  & $0.108$ & $0.160$ & $0.177$ & ($22.1$,$6.0$,$9.8$) & ($3.9$,$15.1$,$2.1$) & \textcolor{white}{+}$0.002$ & $0.013$\\
\hline
Helmi Stream & 5 & $-2.326$ & $+0.365$ & $+0.256$ & ($39.5$,$181.6$,$-59.7$) & ($440.0$,$1366.6$,$735.5$) & $-1.350$ & $0.491$\\
 &  & $0.282$ & $0.247$ & $0.171$ & ($137.9$,$59.7$,$161.3$) & ($76.3$,$113.9$,$27.6$) & \textcolor{white}{+}$0.034$ & $0.026$\\
\hline
\enddata
\end{deluxetable*}

%% file: Tables/interesting_substructure_table.tex
\begin{deluxetable*}{l  l  l  c}
\tablecaption{Associations of Identified DTGs \label{tab:interesting_substructure}}
\tablehead{\colhead{Structure} & \colhead{Reference} & \colhead{Associations} & \colhead{Identified DTGs}}
\startdata
\multirow{5}{*}{MW Substructure} & \multirow{5}{*}{\citet{Naidu2020}} & GSE & 3, 5, 7, 8, 10, 11, 16, 17, 18, 21, 22, 27, 28, 29, 33, 36, 41, 42, 43, 46\\ \cline{3-4}
 &  & MWTD & 2, 12, 14, 15, 19, 20, 24, 25, 40, 44, 45, 49\\ \cline{3-4}
 &  & Thamnos & 26, 51\\ \cline{3-4}
 &  & Helmi Stream & 52\\ \cline{3-4}
 &  & Splashed Disk & 31\\ \cline{1-4}
\multirow{5}{*}{Globular Clusters} & \multirow{5}{*}{\citet{Vasiliev2021}} & Ryu~879~(RLGC~2) & 11, 28, 34\\ \cline{3-4}
 &  & NGC~5139~(oCen) & 35, 50\\ \cline{3-4}
 &  & NGC~6284 & 33\\ \cline{3-4}
 &  & NGC~6397 & 9\\ \cline{3-4}
 &  & UKS~1 & 46\\ \cline{1-4}
\multirow{23}{*}{Previous Groups} & \multirow{7}{*}{\citet{Limberg2021a}} & DTG-30 & 18, 22\\ \cline{3-4}
 &  & DTG-10 & 45\\ \cline{3-4}
 &  & DTG-13 & 4\\ \cline{3-4}
 &  & DTG-22 & 34\\ \cline{3-4}
 &  & DTG-23 & 10\\ \cline{3-4}
 &  & DTG-31 & 51\\ \cline{3-4}
 &  & DTG-38 & 46\\ \cline{2-4}
 & \multirow{6}{*}{\citet{Gudin2021}} & CDTG-14 & 4, 39\\ \cline{3-4}
 &  & CDTG-1 & 22\\ \cline{3-4}
 &  & CDTG-6 & 2\\ \cline{3-4}
 &  & CDTG-8 & 2\\ \cline{3-4}
 &  & CDTG-13 & 18\\ \cline{3-4}
 &  & CDTG-22 & 5\\ \cline{2-4}
 & \multirow{2}{*}{\citet{Cordoni2021}} & Sausage & 3, 4, 7, 8, 10, 11, 17, 18, 27, 28, 33, 42\\ \cline{3-4}
 &  & Sequoia & 34, 51\\ \cline{2-4}
 & \multirow{2}{*}{\citet{Li2019}} & GL-1 & 2, 25\\ \cline{3-4}
 &  & GL-4 & 51\\ \cline{2-4}
 & \multirow{2}{*}{\citet{Monty2020}} & Sausage & 7, 8, 27, 42\\ \cline{3-4}
 &  & SeqG1 & 51\\ \cline{2-4}
 & \multirow{2}{*}{\citet{Yuan2020b}} & DTG-14 & 46\\ \cline{3-4}
 &  & DTG-35 & 6\\ \cline{2-4}
 & \multirow{1}{*}{\citet{Myeong2017}} & Comoving & 7, 8, 11, 28\\ \cline{2-4}
 & \multirow{1}{*}{\citet{Sestito2019}} & P & 12\\ \cline{1-4}
\enddata
\end{deluxetable*}

%% file: Tables/full_data_description_table.tex
\startlongtable
\begin{deluxetable*}{c  l  l  l}
\tablecaption{Description of the Initial Sample from the Best and Brightest Survey \label{tab:full_data_descript}}
\tablehead{\colhead{Column} & \colhead{Field} & \colhead{Unit} & \colhead{Description}}
\startdata
1 & Name & $-$ & The name of the star as given by the Reference\\
2 & Source ID & $-$ & The Gaia EDR3 Source ID of the star\\
3 & SMSS ID & $-$ & The SMSS DR2 Source ID of the star\\
4 & 2MASS ID & $-$ & The 2MASS ID of the star\\
5 & RA & (J2000) & The Right Ascension of the star given in hours:minutes:seconds\\
6 & DEC & (J2000) & The Declination of the star given in degrees:minutes:seconds\\
7 & Telescope & $-$ & The telescope that was used to obtain the spectrum of the star\\
8 & Program ID & $-$ & The program ID the observation was obtained under\\
9 & $E(B - V)$ & $-$ & The reddening value obtained from the n-SSPP output taken from \citet{Schlegel1998}\\
10 & $V_{\rm mag}$ & $-$ & The $V$ magnitude of the star as given by the $V_{\rm mag}$ Reference\\
11 & $B - V$ & $-$ & The $B-V$ color of the star as given by the $V_{\rm mag}$ Reference\\
12 & $J_{\rm mag}$ & $-$ & The $J$ magnitude of the star as given by 2MASS\\
13 & $J - K$ & $-$ & The $J-K$ color of the star as given by 2MASS\\
14 & $G_{\rm mag}$ & $-$ & The Gaia $G$ mean magnitude of the star as given by the Gaia Source ID\\
15 & $G_{\rm BP} - G_{\rm RP}$ & $-$ & The Gaia BP $-$ RP color mean magnitude of the star as given by the Gaia Source ID\\
16 & $V_{\rm mag}$ (Gaia) & $-$ & The $V$ magnitude of the star as determined by the transformations from G mag to V mag using\\
 &  &  & $V$ $=$ $G$ $+$ $0.02704$ $-$ $0.01424*(\rm{BP}-\rm{RP})$ $+$ $0.2156*(\rm{BP}-\rm{RP})^2$ $-$\\
 &  &  & $0.01426(\rm{BP}-\rm{RP})^3$ given by \citet{Riello2021}\\
17 & $J_{\rm mag}$ (Gaia) & $-$ & The J magnitude of the star as determined by the transformations from G mag to J mag using $J$\\
 &  &  & $=$ $G$ $-$ $0.01798$ $-$ $1.389*(\rm{BP}-\rm{RP})$ $+$ $0.09338*(\rm{BP}-\rm{RP})^2$ given by\\
 &  &  & \citet{Riello2021}\\
18 & RV$_{\rm CC}$ & (km s$^{-1}$) & The radial velocity as given by cross-correlation techniques (CC)\\
19 & RV$_{CC, Corrected}$ & (km s$^{-1}$) & The radial velocity for the CC corrected by an offset of 7.276 km/s as compared to radial\\
 &  &  & velocity values in common with RV$_{\rm{Gaia}}$\\
20 & Error & (km s$^{-1}$) & The radial velocity error for the CC\\
21 & RV & (km s$^{-1}$) & The radial velocity as given by RV Reference\\
22 & Error & (km s$^{-1}$) & The radial velocity error as given by RV Reference\\
23 & RV$_{\rm Gaia}$ & (km s$^{-1}$) & The radial velocity as given by the Gaia Source ID\\
24 & Error & (km s$^{-1}$) & The radial velocity error as given by the Gaia Source ID\\
25 & Transits & $-$ & The number of transits that Gaia measured the radial velocity as given by the Gaia Source ID\\
26 & Parallax & (mas) & The parallax as given by the Gaia Source ID\\
27 & Error & (mas) & The parallax error as given by the Gaia Source ID\\
28 & Distance & (kpc) & The inverse parallax distance (1/Parallax)\\
29 & Error & (kpc) & The inverse parallax distance error (Parallax$_{error}$/(Parallax$^2$))\\
30 & Distance$_{\rm Corrected}$ & (kpc) & The corrected inverse parallax distance (1/(Parallax + 0.026)) based on \citet{Huang2021a}\\
31 & Error & (kpc) & The corrected inverse parallax distance error (Parallax$_{error}$/((Parallax + 0.026)$^2$))\\
 &  &  & based on \citet{Huang2021a}\\
32 & Relative Error & $-$ & The relative error of the corrected distance as given by Gaia\\
33 & Distance BJ21 & (kpc) & The 50 percentile distance as given by \citet{Bailer-Jones2021} based on the Gaia Source ID\\
34 & Error & (kpc) & The 50 percentile error as estimated by the 84 percentile distance and the 16 percentile\\
 &  &  & distance as given by \citet{Bailer-Jones2021} based on the Gaia Source ID ((dist84-dist16)/2)\\
35 & Relative Error & $-$ & The relative error of the 50 percentile distance as given by \citet{Bailer-Jones2021} based on\\
 &  &  & the Gaia Source ID\\
36 & PM$_{\rm RA}$ & (mas yr$^{-1}$) & The proper motion in the Right Ascension as given by the Gaia Source ID\\
37 & Error & (mas yr$^{-1}$) & The proper motion error in the Right Ascension as given by the Gaia Source ID\\
38 & PM$_{\rm DEC}$ & (mas yr$^{-1}$) & The proper motion in the Declination as given by the Gaia Source ID\\
39 & Error & (mas yr$^{-1}$) & The proper motion error in the Declination as given by the Gaia Source ID\\
40 & Correlation Coefficient & $-$ & The correlation coefficient between the proper motion in Right Ascension and the proper motion\\
 &  &  & in Declination as given by the Gaia Source ID\\
41 & T$_{\rm eff~Spec}$ & (K) & The effective temperature of the star as given by the n-SSPP\\
42 & Error & (K) & The effective temperature error of the star as given by the n-SSPP\\
43 & T$_{\rm eff~Phot}$ & (K) & The effective temperature of the star as given by \citet{Huang2021c}\\
\pagebreak \\[-3.5ex]
44 & Error & (K) & The effective temperature error of the star as given by \citet{Huang2021c}\\
45 & T$_{\rm eff}$ & (K) & The adopted effective temperature of the star based on the Parameter Procedure\\
46 & Error & (K) & The adopted effective temperature error of the star based on the Parameter Procedure\\
47 & log \textit{g} & (cgs) & The surface gravity of the star as given by the n-SSPP\\
48 & Error & (cgs) & The surface gravity error of the star as given by the n-SSPP\\
49 & [Fe/H]$_{Spec}$ & $-$ & The metallicity of the star as given by the n-SSPP\\
50 & Error & $-$ & The metallicity error of the star as given by the n-SSPP\\
51 & [Fe/H]$_{Phot}$ & $-$ & The metallicity of the star as given by \citet{Huang2021c}\\
52 & Error & $-$ & The metallicity error of the star as given by \citet{Huang2021c}\\
53 & [Fe/H] & $-$ & The adopted metallicity of the star based on the Parameter Procedure\\
54 & Error & $-$ & The adopted metallicity error of the star based on the Parameter Procedure\\
55 & [C/Fe] & $-$ & The carbon abundance ratio for the star\\
56 & Error & $-$ & The carbon abundance ratio error for the star\\
57 & [C/Fe]$_{c}$ & $-$ & The carbon abundance corrected for evolutionary effects from \citet{Placco2014}\\
58 & AC$_{c}$ & $-$ & The absolute carbon corrected for evolutionary effects from \citet{Placco2014} ([C/Fe]$_{c}$ +\\
 &  &  & [Fe/H] + log($\epsilon$)$_{Carbon,Solar}$) (Taken from solar value of 8.43 from\\
 &  &  & \citet{Asplund2009})\\
59 & CARDET & $-$ & Flag with ``D" if the carbon abundance ([C/Fe]) is detected from n-SSPP and ``U" if an upper\\
 &  &  & limit by n-SSPP and ``L" if a lower limit by n-SSPP and ``N" if none is detected by n-SSPP\\
60 & CC$_{\rm [C/Fe]}$ & $-$ & The correlation coefficient of [C/Fe] as given by the n-SSPP\\
61 & CEMP & $-$ & Flag as ``C" for Carbon-Enhanced Metal-Poor (CEMP) if [C/Fe]$_{c}$ $> +0.7$ and ``I" for\\
 &  &  & CEMP-intermediate if $+0.5 <$ [C/Fe]$_{c}$ $\leq +0.7$ and ``N" for Carbon-Normal if\\
 &  &  & [C/Fe]$_{c}$ $\leq +0.5$ and ``X" if there is no [C/Fe]$_{c}$ information\\
62 & [$\alpha$/Fe] & $-$ & The alpha-element abundance ratio for the star\\
63 & Error & $-$ & The alpha-element abundance ratio error for the star\\
64 & ALPDET & $-$ & Flag with ``D" if the alpha abundance ([$\alpha$/Fe]) is detected from n-SSPP and ``U" if an\\
 &  &  & upper limit by n-SSPP and ``L" if a lower limit by n-SSPP and ``N" if none is detected by\\
 &  &  & n-SSPP\\
65 & CC$_{\rm{[}\alpha\rm{/Fe]}}$ & $-$ & The correlation coefficient of [$\alpha$/Fe] as detected by the n-SSPP\\
66 & SNR & $-$ & The average Signal-to-Noise Ratio of the spectrum from n-SSPP\\
67 & Resolving Power & $-$ & The Resolving Power of the spectrum as given by the n-SSPP\\
68 & Reference & $-$ & The Reference for the star as given by ``Placco$\_$2019" for the \citet{Placco2019} sample,\\
 &  &  & ``Schlaufman" for the unpublished Best and Brightest sample and ``SOAR" for the Best and\\
 &  &  & Brightest stars observed with the SOAR telescope and ``Huang" for the photometrically\\
 &  &  & determined Best and Brightest stars from \citet{Huang2021c}\\
69 & Parameter Procedure & $-$ & The procedure used to determine the adopted stellar parameters (T$_{\rm eff}$ and [Fe/H])\\
 &  &  & (``Average" is used if the difference between [Fe/H]$_{Spec}$ and [Fe/H]$_{Phot}$ is less\\
 &  &  & $\leq$\ 0.5 dex and ``Spectroscopic" is used if only [Fe/H]$_{Spec}$ is available and\\
 &  &  & ``Photometric" is used if only [Fe/H]$_{Phot}$ is available, while a choice is made between\\
 &  &  & [Fe/H]$_{Spec}$ and [Fe/H]$_{Phot}$ if both are available and the difference is $>$ 0.5 dex)\\
70 & $V_{\rm mag}$ Reference & $-$ & The Reference for the \textit{V} magnitude of the star\\
71 & Distance AGAMA & $-$ & The Reference for the distance used in AGAMA (BJ21 prioritized over Gaia unless BJ21 distance\\
 &  &  & has relative error greater than 0.3, if both have a relative error greater than 0.3 we adopt\\
 &  &  & no distance estimate)\\
72 & RV Reference & $-$ & The Reference for the RV\\
\enddata
\end{deluxetable*}

%% file: Tables/final_data_description_table.tex
\startlongtable
\begin{deluxetable*}{c  l  l  l}
\tablecaption{Description of the Final Sample from the Best and Brightest Survey \label{tab:final_data_descript}}
\tablehead{\colhead{Column} & \colhead{Field} & \colhead{Unit} & \colhead{Description}}
\startdata
1 & Name & $-$ & The name of the star as given by the Reference\\
2 & Source ID & $-$ & The Gaia EDR3 Source ID of the star\\
3 & SMSS ID & $-$ & The SMSS DR2 Source ID of the star\\
4 & 2MASS ID & $-$ & The 2MASS ID of the star\\
5 & RA & (J2000) & The Right Ascension of the star given in hours:minutes:seconds\\
6 & DEC & (J2000) & The Declination of the star given in degrees:minutes:seconds\\
7 & Telescope & $-$ & The telescope that was used to obtain the spectrum of the star\\
8 & Program ID & $-$ & The program ID the observation was obtained under\\
9 & $E(B - V)$ & $-$ & The reddening value obtained from the n-SSPP output taken from \citet{Schlegel1998}\\
10 & $V_{\rm mag}$ & $-$ & The $V$ magnitude of the star as given by the $V_{\rm mag}$ Reference\\
11 & $B - V$ & $-$ & The $B-V$ color of the star as given by the $V_{\rm mag}$ Reference\\
12 & $J_{\rm mag}$ & $-$ & The $J$ magnitude of the star as given by 2MASS\\
13 & $J - K$ & $-$ & The $J-K$ color of the star as given by 2MASS\\
14 & $G_{\rm mag}$ & $-$ & The Gaia $G$ mean magnitude of the star as given by the Gaia Source ID\\
15 & $G_{\rm BP} - G_{\rm RP}$ & $-$ & The Gaia BP $-$ RP color mean magnitude of the star as given by the Gaia Source ID\\
16 & $V_{\rm mag}$ (Gaia) & $-$ & The $V$ magnitude of the star as determined by the transformations from G mag to V mag using\\
 &  &  & $V$ $=$ $G$ $+$ $0.02704$ $-$ $0.01424*(\rm{BP}-\rm{RP})$ $+$ $0.2156*(\rm{BP}-\rm{RP})^2$ $-$\\
 &  &  & $0.01426(\rm{BP}-\rm{RP})^3$ given by \citet{Riello2021}\\
17 & $J_{\rm mag}$ (Gaia) & $-$ & The J magnitude of the star as determined by the transformations from G mag to J mag using $J$\\
 &  &  & $=$ $G$ $-$ $0.01798$ $-$ $1.389*(\rm{BP}-\rm{RP})$ $+$ $0.09338*(\rm{BP}-\rm{RP})^2$ given by\\
 &  &  & \citet{Riello2021}\\
18 & RV$_{\rm CC}$ & (km s$^{-1}$) & The radial velocity as given by cross-correlation techniques (CC)\\
19 & RV$_{CC, Corrected}$ & (km s$^{-1}$) & The radial velocity for the CC corrected by an offset of 7.276 km/s as compared to radial\\
 &  &  & velocity values in common with RV$_{\rm{Gaia}}$\\
20 & Error & (km s$^{-1}$) & The radial velocity error for the CC\\
21 & RV & (km s$^{-1}$) & The radial velocity as given by RV Reference\\
22 & Error & (km s$^{-1}$) & The radial velocity error as given by RV Reference\\
23 & RV$_{\rm Gaia}$ & (km s$^{-1}$) & The radial velocity as given by the Gaia Source ID\\
24 & Error & (km s$^{-1}$) & The radial velocity error as given by the Gaia Source ID\\
25 & Transits & $-$ & The number of transits that Gaia measured the radial velocity as given by the Gaia Source ID\\
26 & Parallax & (mas) & The parallax as given by the Gaia Source ID\\
27 & Error & (mas) & The parallax error as given by the Gaia Source ID\\
28 & Distance & (kpc) & The inverse parallax distance (1/Parallax)\\
29 & Error & (kpc) & The inverse parallax distance error (Parallax$_{error}$/(Parallax$^2$))\\
30 & Distance$_{\rm Corrected}$ & (kpc) & The corrected inverse parallax distance (1/(Parallax + 0.026)) based on \citet{Huang2021a}\\
31 & Error & (kpc) & The corrected inverse parallax distance error (Parallax$_{error}$/((Parallax + 0.026)$^2$))\\
 &  &  & based on \citet{Huang2021a}\\
32 & Relative Error & $-$ & The relative error of the corrected distance as given by Gaia\\
33 & Distance BJ21 & (kpc) & The 50 percentile distance as given by \citet{Bailer-Jones2021} based on the Gaia Source ID\\
34 & Error & (kpc) & The 50 percentile error as estimated by the 84 percentile distance and the 16 percentile\\
 &  &  & distance as given by \citet{Bailer-Jones2021} based on the Gaia Source ID ((dist84-dist16)/2)\\
35 & Relative Error & $-$ & The relative error of the 50 percentile distance as given by \citet{Bailer-Jones2021} based on\\
 &  &  & the Gaia Source ID\\
36 & PM$_{\rm RA}$ & (mas yr$^{-1}$) & The proper motion in the Right Ascension as given by the Gaia Source ID\\
37 & Error & (mas yr$^{-1}$) & The proper motion error in the Right Ascension as given by the Gaia Source ID\\
38 & PM$_{\rm DEC}$ & (mas yr$^{-1}$) & The proper motion in the Declination as given by the Gaia Source ID\\
39 & Error & (mas yr$^{-1}$) & The proper motion error in the Declination as given by the Gaia Source ID\\
40 & Correlation Coefficient & $-$ & The correlation coefficient between the proper motion in Right Ascension and the proper motion\\
 &  &  & in Declination as given by the Gaia Source ID\\
41 & T$_{\rm eff~Spec}$ & (K) & The effective temperature of the star as given by the n-SSPP\\
42 & Error & (K) & The effective temperature error of the star as given by the n-SSPP\\
43 & T$_{\rm eff~Phot}$ & (K) & The effective temperature of the star as given by \citet{Huang2021c}\\
\pagebreak \\[-3.5ex]
44 & Error & (K) & The effective temperature error of the star as given by \citet{Huang2021c}\\
45 & T$_{\rm eff}$ & (K) & The adopted effective temperature of the star based on the Parameter Procedure\\
46 & Error & (K) & The adopted effective temperature error of the star based on the Parameter Procedure\\
47 & log \textit{g} & (cgs) & The surface gravity of the star as given by the n-SSPP\\
48 & Error & (cgs) & The surface gravity error of the star as given by the n-SSPP\\
49 & [Fe/H]$_{Spec}$ & $-$ & The metallicity of the star as given by the n-SSPP\\
50 & Error & $-$ & The metallicity error of the star as given by the n-SSPP\\
51 & [Fe/H]$_{Phot}$ & $-$ & The metallicity of the star as given by \citet{Huang2021c}\\
52 & Error & $-$ & The metallicity error of the star as given by \citet{Huang2021c}\\
53 & [Fe/H] & $-$ & The adopted metallicity of the star based on the Parameter Procedure\\
54 & Error & $-$ & The adopted metallicity error of the star based on the Parameter Procedure\\
55 & [C/Fe] & $-$ & The carbon abundance ratio for the star\\
56 & Error & $-$ & The carbon abundance ratio error for the star\\
57 & [C/Fe]$_{c}$ & $-$ & The carbon abundance corrected for evolutionary effects from \citet{Placco2014}\\
58 & AC$_{c}$ & $-$ & The absolute carbon corrected for evolutionary effects from \citet{Placco2014} ([C/Fe]$_{c}$ +\\
 &  &  & [Fe/H] + log($\epsilon$)$_{Carbon,Solar}$) (Taken from solar value of 8.43 from\\
 &  &  & \citet{Asplund2009})\\
59 & CARDET & $-$ & Flag with ``D" if the carbon abundance ([C/Fe]) is detected from n-SSPP and ``U" if an upper\\
 &  &  & limit by n-SSPP and ``L" if a lower limit by n-SSPP and ``N" if none is detected by n-SSPP\\
60 & CC$_{\rm [C/Fe]}$ & $-$ & The correlation coefficient of [C/Fe] as given by the n-SSPP\\
61 & CEMP & $-$ & Flag as ``C" for Carbon-Enhanced Metal-Poor (CEMP) if [C/Fe]$_{c}$ $> +0.7$ and ``I" for\\
 &  &  & CEMP-intermediate if $+0.5 <$ [C/Fe]$_{c}$ $\leq +0.7$ and ``N" for Carbon-Normal if\\
 &  &  & [C/Fe]$_{c}$ $\leq +0.5$ and ``X" if there is no [C/Fe]$_{c}$ information\\
62 & [$\alpha$/Fe] & $-$ & The alpha-element abundance ratio for the star\\
63 & Error & $-$ & The alpha-element abundance ratio error for the star\\
64 & ALPDET & $-$ & Flag with ``D" if the alpha abundance ([$\alpha$/Fe]) is detected from n-SSPP and ``U" if an\\
 &  &  & upper limit by n-SSPP and ``L" if a lower limit by n-SSPP and ``N" if none is detected by\\
 &  &  & n-SSPP\\
65 & CC$_{\rm{[}\alpha\rm{/Fe]}}$ & $-$ & The correlation coefficient of [$\alpha$/Fe] as detected by the n-SSPP\\
66 & SNR & $-$ & The average Signal-to-Noise Ratio of the spectrum from n-SSPP\\
67 & Resolving Power & $-$ & The Resolving Power of the spectrum as given by the n-SSPP\\
68 & Reference & $-$ & The Reference for the star as given by ``Placco$\_$2019" for the \citet{Placco2019} sample,\\
 &  &  & ``Schlaufman" for the unpublished Best and Brightest sample and ``SOAR" for the Best and\\
 &  &  & Brightest stars observed with the SOAR telescope and ``Huang" for the photometrically\\
 &  &  & determined Best and Brightest stars from \citet{Huang2021c}\\
69 & Parameter Procedure & $-$ & The procedure used to determine the adopted stellar parameters (T$_{\rm eff}$ and [Fe/H])\\
 &  &  & (``Average" is used if the difference between [Fe/H]$_{Spec}$ and [Fe/H]$_{Phot}$ is less\\
 &  &  & $\leq$\ 0.5 dex and ``Spectroscopic" is used if only [Fe/H]$_{Spec}$ is available and\\
 &  &  & ``Photometric" is used if only [Fe/H]$_{Phot}$ is available, while a choice is made between\\
 &  &  & [Fe/H]$_{Spec}$ and [Fe/H]$_{Phot}$ if both are available and the difference is $>$ 0.5 dex)\\
70 & $V_{\rm mag}$ Reference & $-$ & The Reference for the \textit{V} magnitude of the star\\
71 & Distance AGAMA & $-$ & The Reference for the distance used in AGAMA (BJ21 prioritized over Gaia unless BJ21 distance\\
 &  &  & has relative error greater than 0.3, if both have a relative error greater than 0.3 we adopt\\
 &  &  & no distance estimate)\\
72 & RV Reference & $-$ & The Reference for the RV\\
73 & (v$_{\rm r}$,v$_{\phi}$,v$_{\rm z}$)  & (km s$^{-1}$) & The cylindrical velocities of the star as given by AGAMA\\
74 & Error & (km s$^{-1}$) & The cylindrical velocity errors of the star as given by Monte Carlo sampling through AGAMA\\
75 & (J$_{\rm r}$,J$_{\phi}$,J$_{\rm z}$) & (kpc km s$^{-1}$) & The cylindrical actions of the star as given by AGAMA\\
76 & Error & (kpc km s$^{-1}$) & The cylindrical action errors of the star as given by Monte Carlo sampling through AGAMA\\
77 & Energy & (km$^{2}$ s$^{-2}$) & The orbital energy of the star as given by AGAMA\\
78 & Error  & (km$^{2}$ s$^{-2}$) & The orbital energy error of the star as given by Monte Carlo sampling through AGAMA\\
79 & $r_{\rm peri}$ & (kpc) & The Galactic pericentric distance of the star as given by AGAMA\\
\pagebreak \\[-3.5ex]
80 & Error & (kpc) & The Galactic pericentric distance error of the star as given by Monte Carlo sampling through\\
 &  &  & AGAMA\\
81 & $r_{\rm apo}$ & (kpc) & The Galactic apocentric distance of the star as given by AGAMA\\
82 & Error & (kpc) & The Galactic apocentric distance error of the star as given by Monte Carlo sampling through\\
 &  &  & AGAMA\\
83 & Z$_{\rm max}$ & (kpc) & The maximum height above the Galactic plane of the star as given by AGAMA\\
84 & Error & (kpc) & The maximum height above the Galactic plane error of the star as given by Monte Carlo sampling\\
 &  &  & through AGAMA\\
85 & Eccentricity & $-$ & The eccentricity of the star given by ($r_{\rm{apo}}$ $-$ $r_{\rm{peri}}$)/($r_{\rm{apo}}$ $+$\\
 &  &  & $r_{\rm{peri}}$) through AGAMA\\
86 & Error & $-$ & The eccentricity error of the star as given by Monte Carlo sampling through AGAMA\\
\enddata
\end{deluxetable*}